\documentclass[submission, Phys]{SciPost}

\usepackage{graphicx}
\usepackage{epsfig}
\usepackage[english]{babel}
\usepackage{amsmath}
\usepackage{amssymb}
\usepackage{amsfonts}
\usepackage{bm}
\usepackage{bbm}
\usepackage{fixmath}
\usepackage{bbold}
\usepackage{color,array}
\usepackage{dsfont}

\begin{document}

\newcommand{\red}[1]{\textcolor{black}{#1}}
\newcommand{\blue}[1]{\textcolor{blue}{#1}}

\newcommand{\tr}{\text{Tr}}
\newcommand{\be}{\begin{equation}}
\newcommand{\ee}{\end{equation}}
\newcommand{\bs}{\begin{subequations}}
\newcommand{\es}{\end{subequations}}
\newcommand{\ba}{\begin{align}}
\newcommand{\ket}[1]{{|#1\rangle}}
\newcommand{\bra}[1]{{\langle#1|}}
\newcommand{\trp}{{T_\mathrm{rep}}}
\newcommand{\id}{\mathbb{1}}
\newcommand{\jq}{J_\text{Q}}
\newcommand{\jm}{J_\text{max}}
\newcommand{\am}{A_\text{max}}
\newcommand{\kb}{k_\text{B}}

\begin{center}{\Large \textbf{
Quantum Coherence from Commensurate Driving with Laser Pulses and Decay
}}\end{center}

\begin{center}
G\"otz S.\ Uhrig
\end{center}

\begin{center}
Lehrstuhl f\"{u}r Theoretische Physik I, 
TU Dortmund University\\
Otto-Hahn Stra\ss{}e 4, D-44221 Dortmund, Germany
\\
goetz.uhrig@tu-dortmund.de
\end{center}

\begin{center}
\today
\end{center}


\section*{Abstract}
{\bf
Non-equilibrium physics is a particularly fascinating field of
current research. Generically, driven systems are gradually heated up
so that quantum effects die out. In contrast, we show that
a driven central spin model including
controlled dissipation in a highly excited state allows us
to distill quantum coherent states, indicated by a substantial
reduction of entropy; the key resource is the commensurability
between the periodicity of the pump pulses and the internal processes. 
The model is experimentally accessible
in purified quantum dots or molecules with unpaired electrons.
The potential of preparing and manipulating coherent states
by designed driving potentials is pointed out.
}

\vspace{10pt}
\noindent\rule{\textwidth}{1pt}
\tableofcontents\thispagestyle{fancy}
\noindent\rule{\textwidth}{1pt}
\vspace{10pt}

\section{Introduction}
\label{sec:introduction}

Controlling a quantum mechanical system in a coherent way is one of
the long-standing goals in physics. Obviously, coherent control is 
a major ingredient for handling quantum information. In parallel, 
non-equilibrium physics of quantum systems 
is continuing to attract significant interest.
A key issue in this field is to manipulate systems in time
such that their properties can be tuned and changed
at will. Ideally, they display properties qualitatively different from
what can be observed in equilibrium systems.
These current developments illustrate the interest in 
understanding the dynamics induced by time-dependent Hamiltonians
$H(t)$.

The unitary time evolution operator $U(t_2,t_1)$ induced by $H(t)$ 
is formally given by
\be
U(t_2,t_1) = {\cal T}\exp\left(-i\int_{t_1}^{t_2}H(t)dt\right)
\ee
where ${\cal T}$ is the time ordering operator. While the explicit
calculation of $U(t_2,t_1)$ can be extremely difficult it is obvious
that the dynamics induced by a time-dependent Hamiltonian 
maps quantum states at $t_1$ to quantum states at $t_2$ 
bijectively and conserves the mutual scalar products. Hence, if initially 
the system is in a mixed state with high entropy $S>0$ it stays in a
mixed state for ever with exactly the same entropy.
No coherence can be generated in this way even for a
complete and ideal control of $H(t)$ in time. 
Hence, one has to consider open systems. 

The standard way to generate a single state is to bring the system of interest
into thermal contact with a cold system. Generically,
this is an extremely slow process. The targeted quantum states
have to be ground states of some given system. 
Alternatively, optical pumping in general and laser cooling in particular \cite{phill98} are well 
established techniques to lower the entropy of microscopic systems
using resonant pumping and spontaneous decay. Quite recently, engineered
dissipation has been recognized as a means to generate
targeted entangled quantum states in small \red{\cite{witth08,verst09,vollb11}} 
and extended systems \cite{kraus08,diehl08}. Experimentally, entanglement 
has been shown for two quantum bits 
\cite{lin13,shank13} and for two trapped mesoscopic cesium clouds \cite{kraut11}.

In this article, we show that periodic driving  can have a quantum 
system converge to coherent quantum states if an intermediate, highly 
excited  and decaying state is involved. The key aspect is
the commensurability of the \red{period of the pump pulses to the 
time constants of the internal processes, here Larmor precessions}. 
This distinguishes our proposal from established optical pumping
protocols. The completely disordered initial mixture can 
be made almost coherent. The final mixture only
has an entropy $S\approx \kb\ln2$ corresponding to a mixture of
 two states.
An appealing asset is that once the driving is switched off 
the Lindbladian decay does not matter anymore
and the system is governed by Hamiltonian dynamics only.

The focus of the present work
is to exemplarily demonstrate the substantial reduction of entropy in 
a small spin system
subject to periodic laser pulses. The choice of system is motivated
by experiments on the electronic spin in quantum dots interacting with
nuclear spins 
\cite{greil06a,greil07b,petro12,econo14,beuge16,jasch17,scher18,klein18}. 
The model studied is also applicable
to the electronic spin in molecular radicals \cite{bunce87} 
or to molecular magnets, see Refs.\ \cite{blund07,ferra17,schna19}. 
In organic molecules the spin
bath is given by the nuclear spins of the hydrogen nuclei 
in organic ligands.

\section{Model}
\label{sec:model}

The model comprises a central, electronic spin $S=1/2$ which is
coupled to nuclear spins
\be
\label{eq:hamil_spin}
H_\text{spin} = H_\text{CS} + H_\text{eZ} + H_\text{nZ}
\ee
where $H_\text{eZ}=h S^x$ is the electronic Zeeman term with
$h=g\mu_\text{B} B$ ($\hbar$ is set to unity 
\red{here and henceforth) with the gyromagnetic factor
$g$, the Bohr magneton $\mu_\text{B}$,
 the external magnetic field $B$ in $x$-direction and the $x$-component $S^x$
of the central spin. The nuclear Zeeman term is given by 
$H_\text{nZ} = z h \sum_{i=1}^N I^x_i$
where $z$ is the ratio of the nuclear $g$-factor multiplied by 
the nuclear magneton and their electronic counterparts 
$g_\text{nuclear}\mu_\text{nuclear}/(g\mu_\text{B})$.
The operator $I^x_i$ is the $x$-component of the nuclear spin $i$.
For simplicity we take  $I=1/2$ for all nuclear spins.} 
Due to the large nuclear mass, the factor $z$ is of the
order of $10^{-3}$, but in principle other $z$-values can be studied as well,
\red{see also below}.
In the central spin part $H_\text{CS}=\vec{S}\cdot\vec{A}$ the 
\red{so-called Overhauser field $\vec{A}$  results from 
the combined effect of all nuclear spins each of which is
interacting via the hyperfine coupling $J_i$ with} the central spin
\be
\vec{A} = \sum_{i=1}^N J_i \vec{I}_i.
\ee
\red{If the central spin results from an electron the
hyperfine coupling is a contact interaction at the location
of the nucleus stemming from relativistic corrections to the
non-relativistic Schr\"odinger equation with a Coulomb potential. 
It is proportional to the
probability of the electron to be at the nucleus, i.e., to the
modulus squared of the electronic wave function \cite{schli03,coish04}. Depending on 
the positions of the nuclei and on the shape of the wave function
various distributions of the $J_i$ are plausible. A 
Gaussian wave function in one dimension implies
a parametrization by a Gaussian while in two dimensions
an exponential parametrization is appropriate \cite{farib13a,fause17a}
distribution. We will first use a uniform distribution for simplicity
and consider the Gaussian and exponential case afterwards.}

Besides the spin system there is an important intermediate state
given by a single trion state $\ket{\mathrm{T}}$ \red{consisting of the
single fermion providing the central spin bound to an
additional exciton. This trion is polarised in $z$-direction
at the very high energy $\varepsilon$ ($\approx 1$ eV).
The other polarisation exists as well, but using circularly 
polarised light it is not excited. A Larmor precession of
the trion is not considered here for simplicity.
Then,} the total Hamiltonian reads
\be
H = H_\text{spin} + \varepsilon \ket{\mathrm{T}}\bra{\mathrm{T}}.
\ee
The laser pulse is taken to be very short as in experiment
where its duration $\tau$ is of the order of picoseconds. Hence, we
describe \red{its effect by a unitary time evolution operator
 $\exp(-i\tau H_\text{puls})=U_\text{puls}$} which excites the $\ket{\uparrow}$
state of the central spin to the trion state or de-excites it 
\be
\label{eq:puls}
U_\text{puls} = c^\dag + c +\ket{\downarrow} \bra{\downarrow}.
\ee
where $c:=\ket{\uparrow}\bra{\mathrm{T}}$ and
$c^\dagger:=\ket{\mathrm{T}}\bra{\uparrow}$. 
\red{This unitary operator happens to be hermitian as well, but this
is not an important feature. 
One easily verifies $U_\text{puls}U_\text{puls}^\dag=\mathbb{1}$.}
 Such pulses are applied in long periodic trains lasting seconds and minutes. 
The repetition time between two consecutive pulses is $\trp$ of the 
order of 10 ns.

The decay of the trion is described by the Lindblad equation
for the density matrix $\rho$
\be
\label{eq:lind}
\partial_t \rho(t) = -i[H,\rho] - \gamma (c^\dag c\rho + \rho c^\dag c- 2c\rho c^\dag)
\ee
where the prefactor $\gamma>0$ of the dissipator term \cite{breue06}
defines the decay rate. The corresponding process with $c$ and $c^\dag$
swapped needs not be included because its decay rate is smaller
by $\exp(-\beta\varepsilon)$, i.e., it vanishes for all physical purposes.
\red{We emphasize that we deal with an open quantum system by virtue of 
the Lindblad dynamics in \eqref{eq:lind}.
Since the decay of the trion generically implies the emission of
a photon at high energies 
the preconditions for using Lindblad dynamics are perfectly met \cite{breue06}.}

\section{Mathematical Properties of Time Evolution}
\label{sec:math-proper}

The key observation is that the dynamics from just before the $n$th pulse
at $t=n\trp -$ to just before the $n+1$st pulse at $t=(n+1)\trp -$
is a \emph{linear} mapping $M: \rho(n\trp-) \rightarrow \rho((n+1)\trp-)$ which
 does not depend on $n$. Since it is 
acting on operators one may call it a superoperator.
Its matrix form is derived explicitly in Appendix \ref{app:matrix}. 
If no dissipation took place ($\gamma=0$) the
mapping $M$ would be unitary. But in presence of the dissipative trion decay
it is a general matrix with the following properties:
\begin{enumerate}
\item
The matrix $M$ has an eigenvalue $1$ which may be degenerate.
If the dynamics of the system takes place in $n$ separate
subspaces without transitions between them the degeneracy
is at least $n$.
\item
All eigenoperators to eigenvalues different from 1 are traceless.
\item
At least one eigenoperator to eigenvalue 1 has a finite trace.
\item
The absolute values of all eigenvalues of $M$ are not larger
than 1. 
\item
If there is a non-real eigenvalue $\lambda$ with eigenoperator $C$,
the complex conjugate $\lambda^*$ is also an eigenvalue with eigenoperator
$C^\dag$.
\item
The eigenoperators to eigenvalues 1 can be scaled to be
hermitian. 
\end{enumerate}
While the above properties can be shown rigorously, see Appendix \ref{app:properties},
for any Lindblad evolution,
the following ones are observed numerically in the analysis of the  
particular model \eqref{eq:lind} under study here:
\begin{itemize}
\item[(a)]
The matrix $M$ is diagonalizable; it does not require a Jordan normal form.
\item[(b)]
For pairwise different couplings $i\ne j\Rightarrow J_i\ne J_j$ the
eigenvalue $1$ is non-degenerate.
\item[(c)]
The eigenoperators to eigenvalue 1 can be scaled to be hermitian and non-negative.
In the generic, non-degenerate case we denote the properly scaled eigenoperator 
$V_0$ with $\tr(V_0)=1$.
\item[(d)]
No eigenvalue $\neq 1$, but with absolute value 1, occurs,
i.e., all eigenvalues different from 1 are smaller than 1 in absolute value.
\item[(e)]
Complex eigenvalues and complex eigenoperators do occur.
\end{itemize}

The above properties allow us to understand what happens in experiment
upon application of long trains of pulses corresponding
to $10^{10}$ and more applications of $M$. Then it is safe to conclude that all
contributions from eigenoperators to eigenvalues smaller than 1 have died out
completely. Only the (generically) single eigenoperator $V_0$ to eigenvalue 1
is left such that
\be
\lim_{n\to\infty} \rho(n\trp-) = V_0.
\ee
The quasi-stationary state after long trains of pulses is given by $V_0$
\footnote{We use the term `quasi-stationary' state
because it is stationary only if we detect it stroboscopically at the
time instants $t=n\trp-$.}.
This observation simplifies the calculation of the long-time limit
 greatly compared to previous quantum mechanical studies 
\cite{econo14,beuge16,beuge17,klein18}. 
One has to compute the eigenoperator of $M$ to 
the eigenvalue 1. Below this is performed by diagonalization of $M$ which 
is a reliable approach, but restricted to small systems $N\lessapprox6$.
We stress that no complete diagonalization is required to know $V_0$ 
because only the eigenoperator to the eigenvalue 1 is needed.
Hence we are optimistic that further computational improvements are possible.
If, however, the speed of convergence is of interest more information on the
spectrum and the eigenoperators of $M$ is needed, see also Sect.\ \ref{sec:convergence}.


\section{Results on Entropy}
\label{sec:entropy}

It is known that in pulsed quantum dots
nuclear frequency focusing occurs (NFF) \cite{greil06a,greil07b,evers18} 
which can be explained by a 
significant change in the distribution of the Overhauser field
\cite{petro12,econo14,beuge16,beuge17,scher18,klein18} which is Gaussian
initially. This distribution
develops a comb structure with equidistant spikes. The difference $\Delta A_x$ 
between consecutive spikes is such that it corresponds to a full additional revolution
of the central spin $\trp \Delta A_x=2\pi$. A comb-like
probability distribution is more structured and contains more information 
than the initial featureless Gaussian. For instance, the entropy reduction
 of the Overhauser field distributions computed
in Ref.\ \cite{klein18}, Fig.\ 12, relative to the initial Gaussians is 
$\Delta S =-0.202\kb$ at $B=0.93$T and $\Delta S =-0.018\kb$ at $B=3.71$T.
Hence, NFF decreases the entropy, but only slightly for large spin
baths. This observation inspires us to
ask to which extent continued pulsing can reduce entropy
and which characteristics the final state has.

Inspired by the laser experiments on quantum dots \cite{greil06a,greil07b,evers18}
we choose an (arbitrary) energy unit $J_\text{Q}$ and thus $1/J_\text{Q}$, \red{recalling
that we have set $\hbar=1$}, 
as time unit which can be assumed to be of the order of 1ns. The repetition time 
$\trp$ is set to $4\pi/J_\text{Q}$ which is
on the one hand close to the experimental values where $\trp=13.2\text{ns}$ and
 on the other hand makes it easy to recognize
resonances, see below. The trion decay rate is set to $2\gamma=2.5 J_\text{Q}$
to reflect a trion life time of $\approx 0.4\red{\text{ns}}$. The bath size is restricted
to $N\in\{1,2,\ldots,6\}$, but still allows us to draw fundamental conclusions
and to describe electronic spins coupled to hydrogen nuclear spins in small
molecules \cite{bunce87,blund07,ferra17,schna19}.
The individual couplings $J_i$ are chosen to be distributed according to 
\be
\label{eq:equidistant}
J_i = J_\text{max}(\sqrt{5}-2)\left(\sqrt{5}+2({i-1})/({N-1})\right),
\ee
which is a uniform distribution between $J_\text{min}$ and $J_\text{max}$ with
$\sqrt{5}$ inserted to avoid accidental commensurabilities \red{of the
different couplings $J_i$}. \red{The value $J_\text{min}$
results from $J_i$ for $i=1$.
Other parametrizations are motivated by the shape of the electronic wave
functions \cite{merku02,schli03,coish04}.}
Results for a frequently used exponential parameterization \cite{farib13a}
\be
\label{eq:expo}
J_i = J_\text{max}\exp(-\alpha(i-1)/(N-1))
\ee
with $\alpha\in\{0.5, 1\}$ and for a Gaussian parametrization,
motivated by the electronic wave function in quantum dots \cite{coish04},
\be
\label{eq:gaus}
J_i = J_\text{max}\exp(-\alpha[(i-1)/(N-1)]^2).
\ee
are given in the next section and in Appendix \ref{app:other}.
\red{For both parametrizations the minimum value $J_\text{min}$ occurs for $i=N$
and takes the value $J_\text{min}=J_\text{max}\exp(-\alpha)$.}

\begin{figure}[htb]
\centering
\includegraphics[width=0.60\columnwidth]{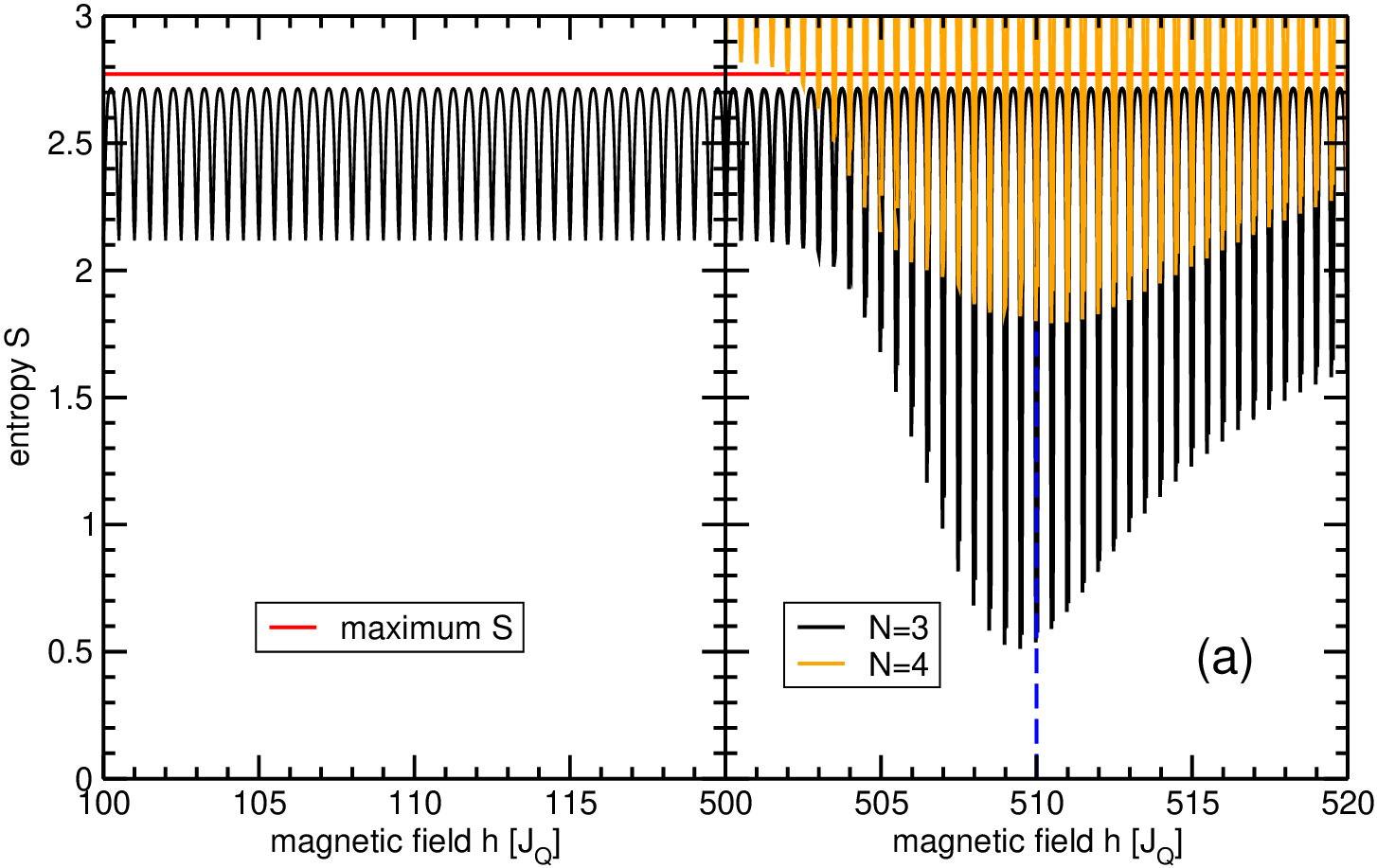} 
\includegraphics[width=0.59\columnwidth]{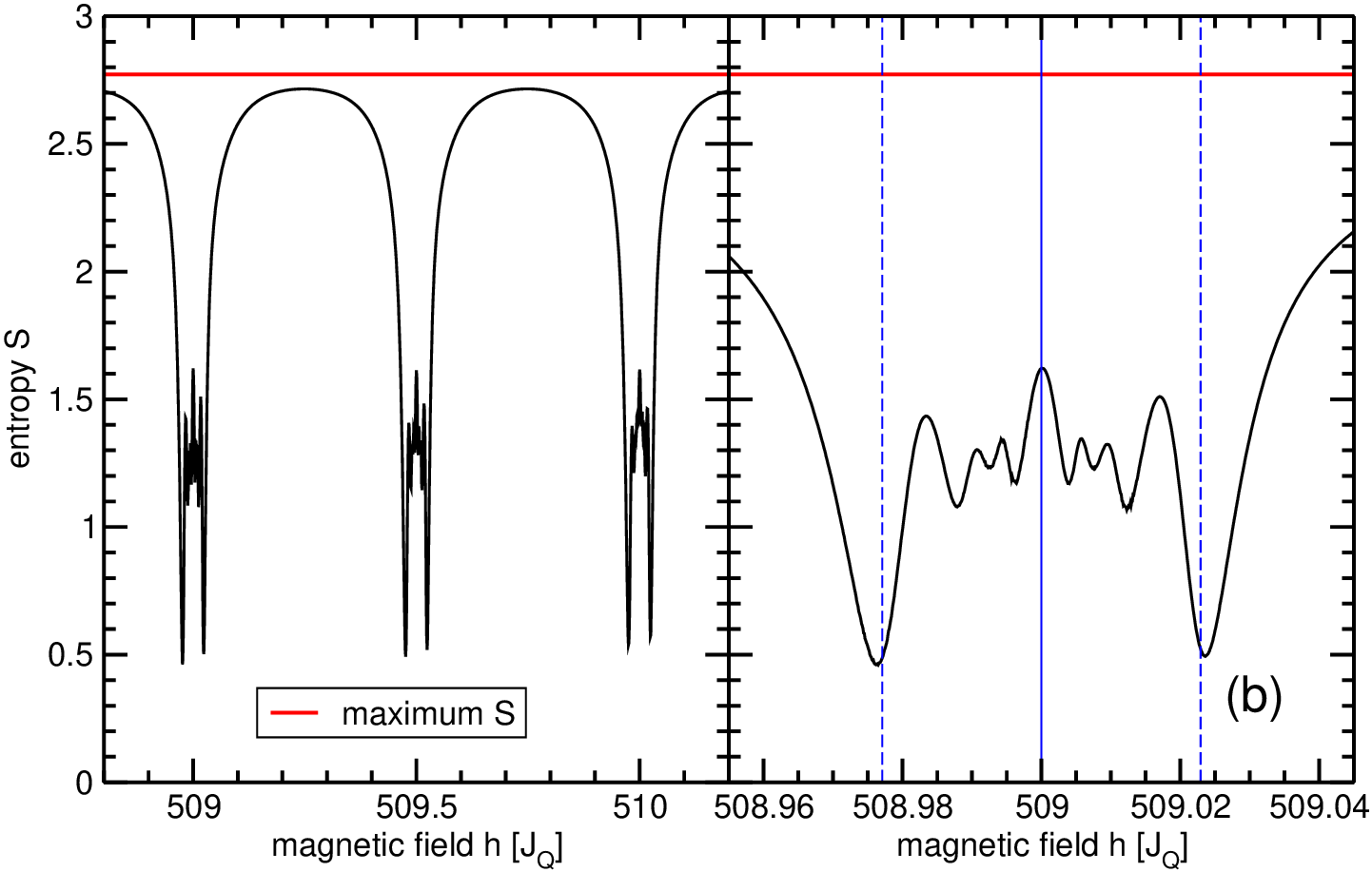} 
\caption{(a) Residual entropy of the limiting density matrix $V_0$ obtained after
infinite number of pulses vs.\ the applied magnetic field for
$\jm=0.02\jq$ and $z=1/1000$; 1 Tesla corresponds roughly to $50\jq$. 
Resonances of the electronic spin occur every
$\Delta h=0.5\jq$; resonances of the nuclear spins occur every $\Delta h = 500\jq$.
The blue dashed line depicts an offset of $\Delta h=\pm\jm/(2z)$ from the nuclear resonance.
(b) Zooms into intervals of the magnetic field where the lowest entropies
are reached. The blue dashed lines depict an offset of $\Delta h =\pm \am$ from the
electronic resonance.}
\label{fig:overview}
\end{figure}

Figure \ref{fig:overview} displays a generic dependence on the external
magnetic field $h=g\mu_\text{B}B_x$ of the entropy of the limiting density
matrix $V_0$ obtained after infinite number of pulses. 
Two nested resonances of the Larmor precessions
are discernible: the central electronic spin resonates for 
\be
\label{eq:res-central}
h\trp = 2\pi n, \qquad n\in\mathbb{Z}
\ee
\red{where $n$ is the number of Larmor revolutions that fit
into the interval $\trp$ between two pulses. This means that for an
increase of the magnetic field from $h$
to $h+\Delta h$ with $\Delta h=2\pi/\trp$ the 
central spin is in the same state before the pulse as it was
at $h$.}

\red{The other resonance is related to the Larmor precession of 
the nuclear bath spins 
which leads to the condition  
\be
\label{eq:res-bath}
zh\trp=2\pi n', \qquad n'\in\mathbb{Z}
\ee
where $n'$ indicates the number of Larmor revolutions of the nuclear spins
which fit between two pulses. Upon increasing the magnetic field $h$,
the nuclear spins are in the same state before the next pulse 
if $h$ is changed to $h+\Delta h$ with $\Delta h=2\pi/(z\trp)$.}

\red{But the two resonance conditions \eqref{eq:res-central} and \eqref{eq:res-bath}
for the central spin and for the bath spins apply precisely as given only without 
coupling between the spins. The coupled system displays important shifts. The 
nuclear resonance appears to be shifted by $z\Delta h \approx \pm \jm/2$,
see right panel of Fig.\ \ref{fig:overview}(a). The explanation is that the dynamics of
the central spin $S=1/2$ creates an additional magnetic field 
\red{similar to a Knight shift} acting on each nuclear spin of the
order of $J_i/2$ which is estimated by $\jm/2$. Further support \red{of the
validity of this} explanation is given in Appendix \ref{app:shift}.}

The electronic resonance is shifted by 
\be
\label{eq:over-shift}
\Delta h = \pm\am
\ee
where $\am$ is the \red{maximum possible value of the}
Overhauser field given by $\am:=(1/2)\sum_{i=1}^N J_i$
for maximally polarized bath spins. This is shown in the right
panel of Fig.\ \ref{fig:overview}(b).

\red{Fig.\ \ref{fig:overview} shows that the effect of the
periodic driving on the 
entropy strongly depends on the precise value of the magnetic field. 
The entropy reduction is largest \emph{close} to the central resonance \eqref{eq:res-central}
and to the bath resonance \eqref{eq:res-bath}. This requires that both
resonances must be approximately commensurate. In addition, the \emph{precise} position of
the maximum entropy reduction depends on the two above shifts, the approximate Knight shift and
the shift by the maximum Overhauser field \eqref{eq:over-shift}.}

We pose the question to 
which extent the initial entropy of complete disorder
$S_\text{init}=\kb(N+1)\ln2$ (in the figures and henceforth
$\kb$ is set to unity) can be reduced by commensurate
periodic pumping.
The results in Fig.\ \ref{fig:overview} clearly show that remarkably low
values of entropy can be reached. The residual value of $S\approx 0.5\kb$
in the minima of the right panel of Fig.\ \ref{fig:overview}(b) corresponds
to a contribution of less than two states ($S=\ln2\kb \approx 0.7\kb$) while initially
16 states were mixed for $N=3$ so that the initial entropy is
$S_\text{init}=4\ln2\kb\approx2.77\kb$. 
This represents a remarkable distillation of coherence.

\begin{figure}[htb]
\centering
\includegraphics[width=0.60\columnwidth]{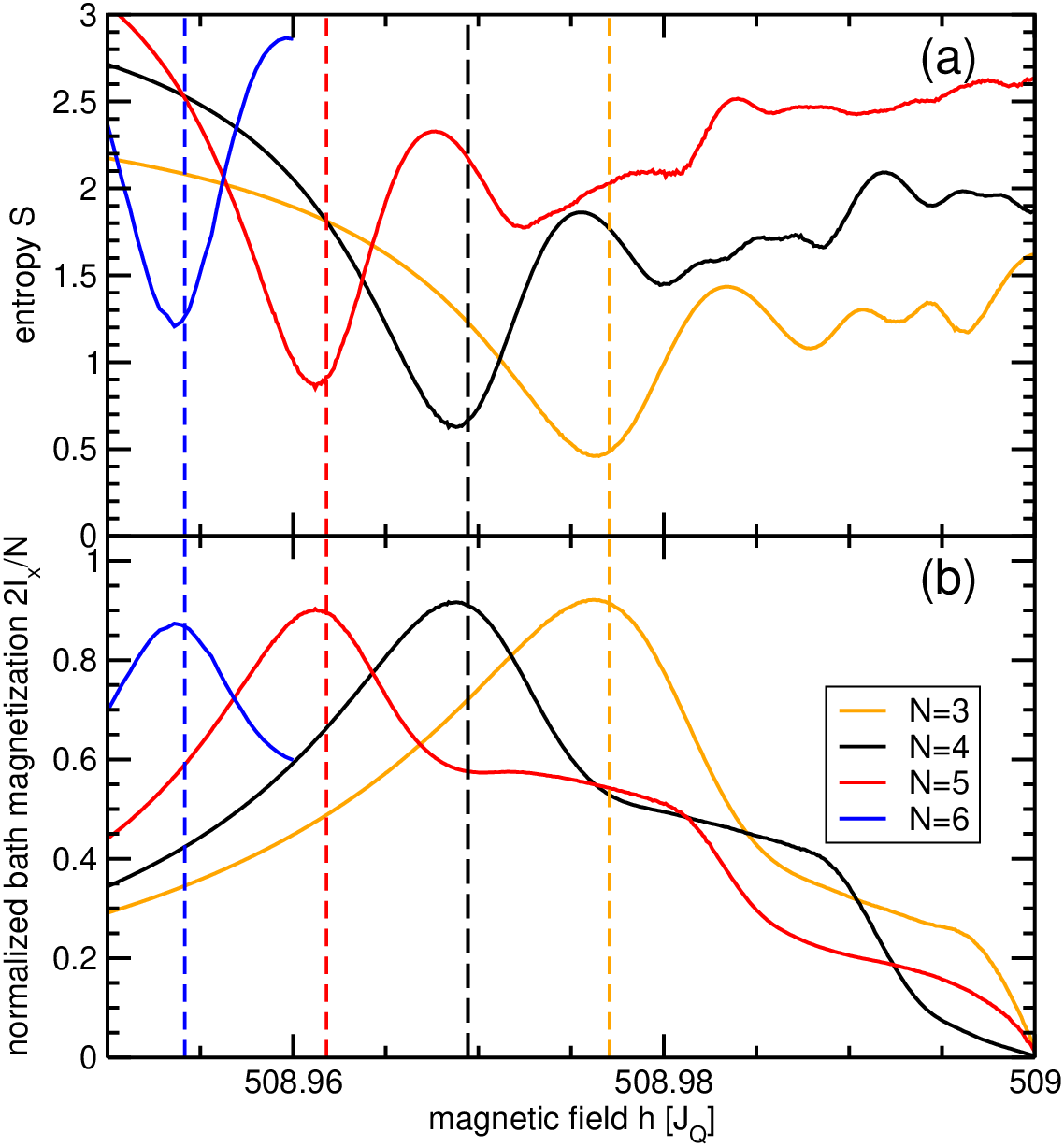} 
\caption{(a) Residual entropy of the limiting density matrix $V_0$ 
for various bath sizes; other parameters as in Fig.\ \ref{fig:overview}. 
The dashed lines indicate the shifts of the electronic
resonance by $-\am$. (b) Corresponding normalized polarization of the spin bath
in the external field direction, i.e.\ the $x$-direction.}
\label{fig:size-mag}
\end{figure}

Hence, we focus on the minima and in particular on the
left minimum. We address the question whether the 
distillation of coherence still works for larger systems.
Unfortunately, 
the numerical analysis cannot be extended easily due to the
dramatically increasing dimension $D= 2^{2(N+1)}$ because
we are dealing with the Hilbert space of density matrices of the spin bath
and the central spin. Yet a 
trend can be deduced from results up to $N=6$ displayed in
Fig.\ \ref{fig:size-mag}(a). The entropy reduction per $N+1$ spins is
$-0.58\kb$ for $N=3$,  $-0.57\kb$ for $N=4$,  $-0.55\kb$ for $N=5$,
and  $-0.52\kb$ for $N=6$. The reduction is substantial, but
slowly decreases with system size. Presently, we cannot know the behavior for
$N\to\infty$. The finite value $\approx-0.2\kb$ found in the semiclassical
simulation \cite{scher18,klein18} indicates that the effect persists for
large baths. In Appendix \ref{app:other}, 
results for the couplings defined in \eqref{eq:expo} or in \eqref{eq:gaus} 
are given which corroborate our finding. The couplings may be rather close to each
other, but not equal. It appears favorable that the spread of couplings
is not too large.

Which state is reached in the minimum
of the residual entropy? The decisive clue is provided by the lower panel
Fig.\ \ref{fig:size-mag}(b) displaying the polarization of the spin bath.
It is normalized such that its saturation value is unity.
Clearly, the minimum of the residual entropy coincides with the maximum
of the polarization. The latter is close to its saturation value though
not quite with a minute decrease for increasing $N$. This tells us that
the limiting density matrix $V_0$ essentially corresponds to the polarized
spin bath. The central electronic spin is also almost perfectly polarized 
(not shown), but in $z$-direction. These observations clarify the 
state which can be retrieved by long trains of pulses.

Additionally, Fig.\ \ref{fig:size-mag}(b) explains the shift of the
electronic resonance. The polarized spin bath renormalizes the external
magnetic field by (almost) $\pm\am$. To the left of the resonance, it enhances
the external field ($+\am$) while the external field is effectively reduced
($-\am$) to the right of the resonance. Note that an analogous direct explanation
for the shift of the nuclear resonance in the right panel of Fig.\
\ref{fig:overview} is not valid. The computed polarization of the
central spin points in $z$-direction and thus does not shift the external
field.

\section{Results on Convergence}
\label{sec:convergence}

In order to assess the speed of convergence of the initially disordered density
matrix $\rho_0=\mathbb{1}/Z$ to the limiting density matrix $V_0$ we
proceed as follows. Let us assume that the matrices $v_i$ are the eigen
matrices of $M$ and that they are normalized $||v_i||^2:=\tr(v_i^\dag v_i)=1$.
Since the mapping $M$ is not unitary, orthogonality of the eigenmatrices cannot
be assumed. Note that the standard normalization generically implies 
that there is some factor between $V_0$ with $\tr(V_0)=1$ and $v_0$.
The initial density matrix $\rho_0$ can be expanded in the $\{v_i\}$
\be
\rho_0 = \sum_{j=0}^{D-1} \alpha_j v_j.
\ee
After $n$ pulses, the density matrix $\rho_n$ is given by
\be
\rho_n = \sum_{j=0}^{D-1} \alpha_j  \lambda_j^n v_j
\ee
where $\lambda_j$ are the corresponding eigenvalues of $M$
and $\lambda_0=1$ by construction.
We aim at $\rho_0$ being close to $V_0$ within $p_\text{thresh}$, i.e.,
\be
\label{eq:converged}
|| \rho_n -V_0 ||\le p_\text{thresh} ||V_0||
\ee
should hold for an appropriate $n$. A generic value of the threshold
$p_\text{thresh}$ is $1\%$. To this end, the minimum $n$ which
fulfills \eqref{eq:converged} has to be estimated.

\begin{figure}[htb]
\centering
\includegraphics[width=0.60\columnwidth]{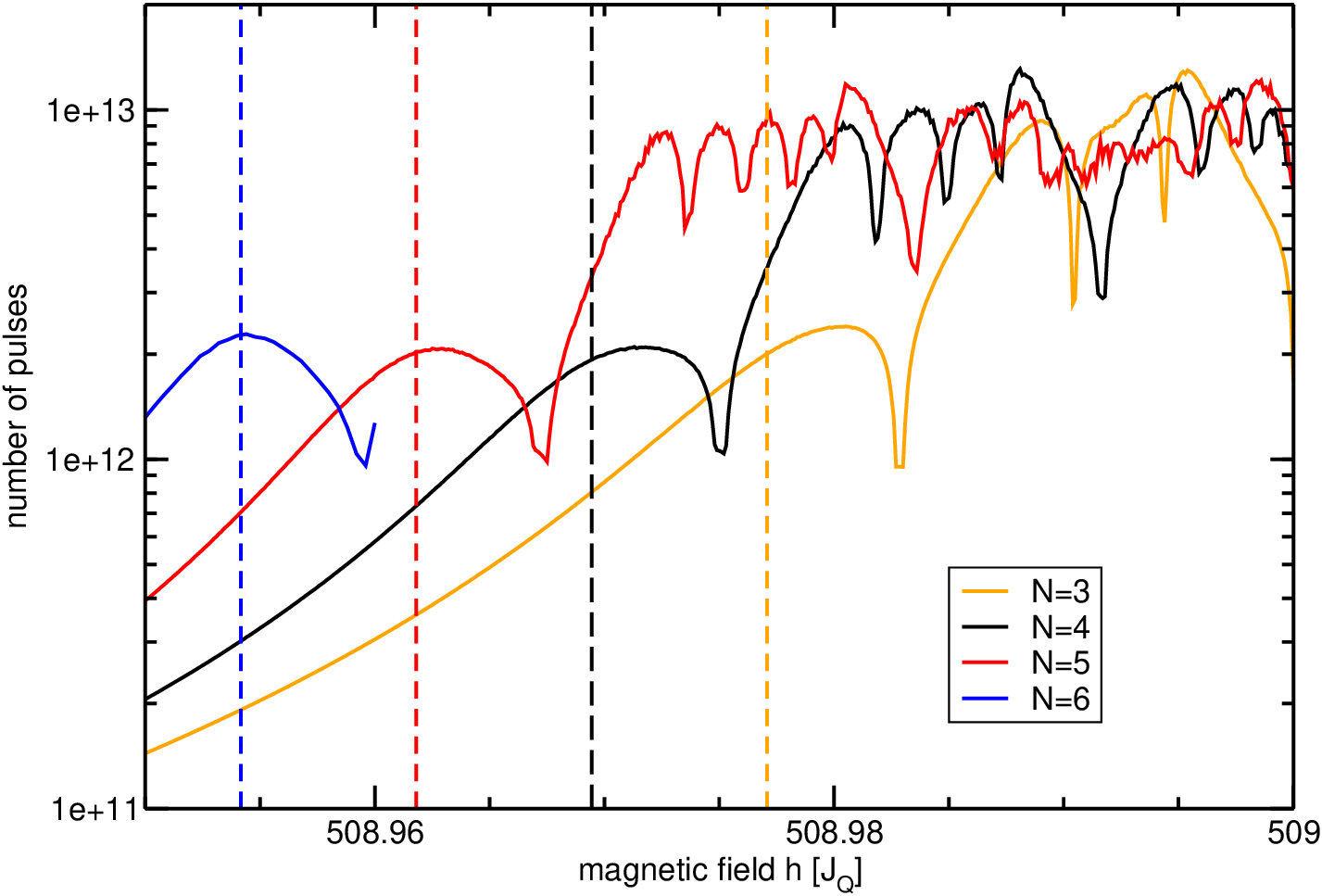} 
\caption{Number of pulses for a convergence within $1\%$ ($p_\text{thresh}=0.0$)
are plotted for various bath sizes; couplings given by \eqref{eq:equidistant},
other parameters as in Fig.\ \ref{fig:overview}.
The corresponding residual entropies and magnetizations 
are depicted in Fig.\ \ref{fig:size-mag}. The vertical dashed lines indicate
the estimates \eqref{eq:over-shift} for the entropy minima as before.}
\label{fig:number-N}
\end{figure}

Such an estimate can be obtained by determining
\be
n_j := 1+\text{trunc}\left[\frac{\ln(|p_\text{thresh}\alpha_0/\alpha_j|)}{\ln(|\lambda_j|)}\right]
\ee
for $j\in\{1, 2,3, \ldots,D-1 \}$. The estimate of the required number of
pulses is the maximum of these number, i.e.,
\be
n_\text{puls} := \max_{1\le j < D} n_j.
\ee
We checked exemplarily that the number determined in this way implies that 
the convergence condition \eqref{eq:converged} is fulfilled. 
This is not mathematically rigorous because it could be that there are very many
slowly decreasing contributions which add up to a significant deviation from
$V_0$. But generically, this is not the case.

In Fig.\ \ref{fig:number-N} the results are shown for various bath sizes
and the parameters for which the data of the previous figures was computed.
Since the entropy minima are located at the positions of the vertical dashed lines
to good accuracy one can read off the required number of pulses at the intersections
of the solid and the dashed lines. Clearly, about $2\cdot 10^{12}$ pulses are necessary
to approach the limiting, relatively pure density matrices $V_0$. 
Interestingly, the number of required pulses does not depend
much on the bath size, at least for the accessible bath sizes.
This is a positive message in view of the scaling towards larger baths
in experimental setups.

\begin{figure}[htb]
\centering
\includegraphics[width=0.60\columnwidth]{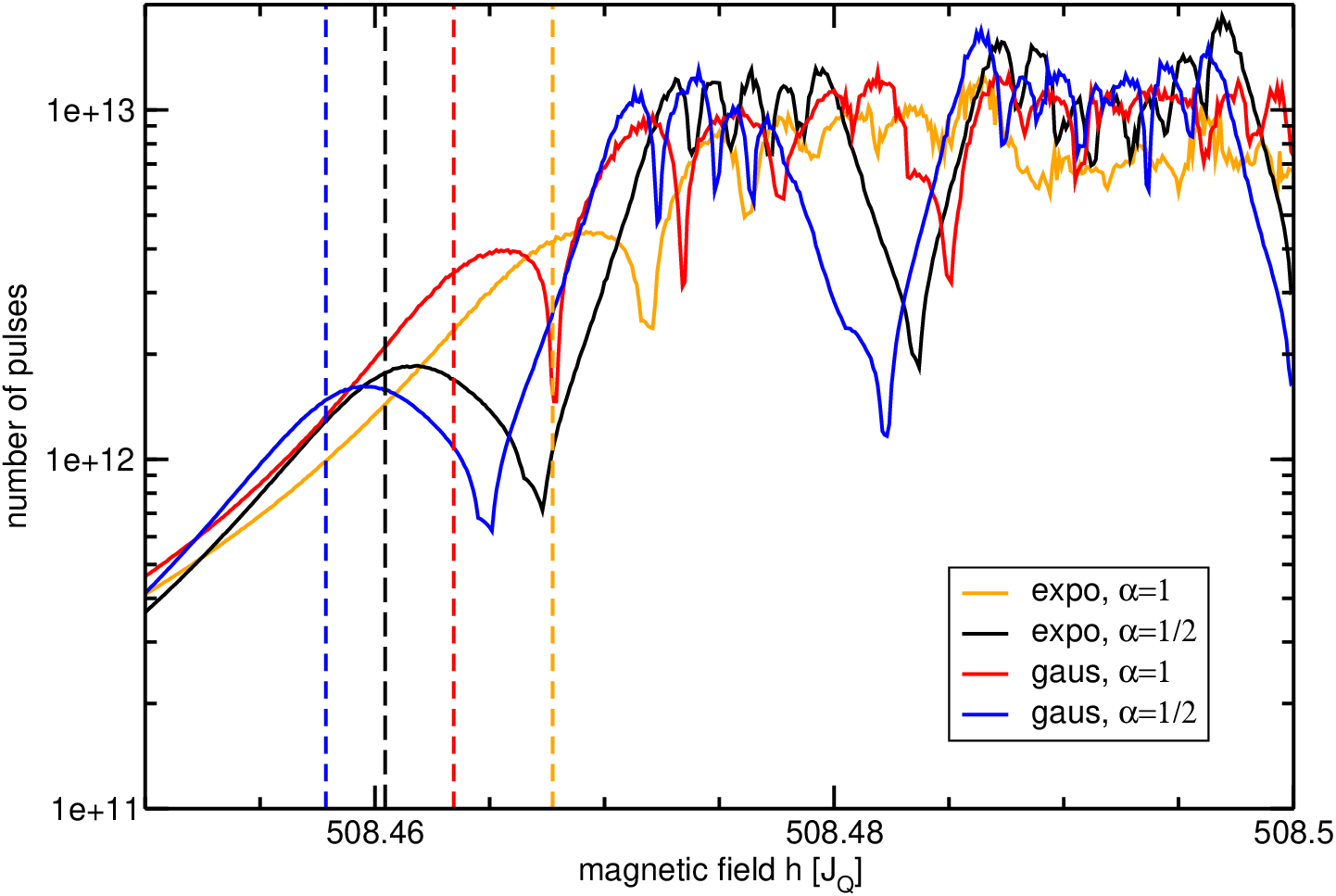} 
\caption{Number of pulses for a convergence within $1\%$ ($p_\text{thresh}=0.01$)
for $N=5$, $\jm=0.02\jq$, and $z=10^{-3}$ for the exponential parametrization
 in \eqref{eq:expo} (legend ``expo'') and the Gaussian parametrization
in \eqref{eq:gaus} (legend ``gaus'').
The corresponding residual entropies and magnetizations 
are depicted in Figs.\ \ref{fig:expo} and \ref{fig:gaus}, respectively. 
The vertical dashed lines indicate
the estimates for the entropy minima which are shifted from the
resonances without interactions according to \eqref{eq:over-shift}.}
\label{fig:parametrization}
\end{figure}

Figure \ref{fig:parametrization} depicts the required minimum number
of pulses for the two alternative parametrizations of the couplings
\eqref{eq:expo} and \eqref{eq:gaus}. Again, the range is about $3\cdot 10^{12}$.
Still, there are relevant differences. The value $n_\text{puls}$ is higher 
for $\alpha=1$ ($\approx 4\cdot 10^{12}$) than for $\alpha=1/2$ ($\lessapprox 2\cdot 10^{12}$).
This indicates that the mechanism of distilling quantum states by 
commensurability with periodic external pulses works best if the couplings \red{$J_i$
are similar, i.e., if their spread given by $J_\text{min}/J_\text{max}=\exp(-\alpha)$} is small. The same qualitative result is obtained for the residual entropy, 
see Appendix  \ref{app:other}.

Note that this argument also explains why the Gaussian parametrized couplings \eqref{eq:gaus}
require slightly less pulses than the exponential parametrized couplings \eqref{eq:expo}.
\red{The couplings $J_i$ cumulate at their maximum $\jm$ in the Gaussian case so that their
variance is slightly smaller than the one of the exponential parametrization.}
One could have thought that 
the cumulated couplings $J_i \approx \jm$ in the Gaussian case 
require longer pulsing in order to achieve a given degree of distillation
because mathematically equal couplings $J_i=J_{i'}$ imply degeneracies preventing
distillation, see the mathematical properties discussed in Sect.\ \ref{sec:math-proper}.
\red{But this appears  not to be the case.}

\begin{figure}[htb]
\centering
\includegraphics[width=0.60\columnwidth]{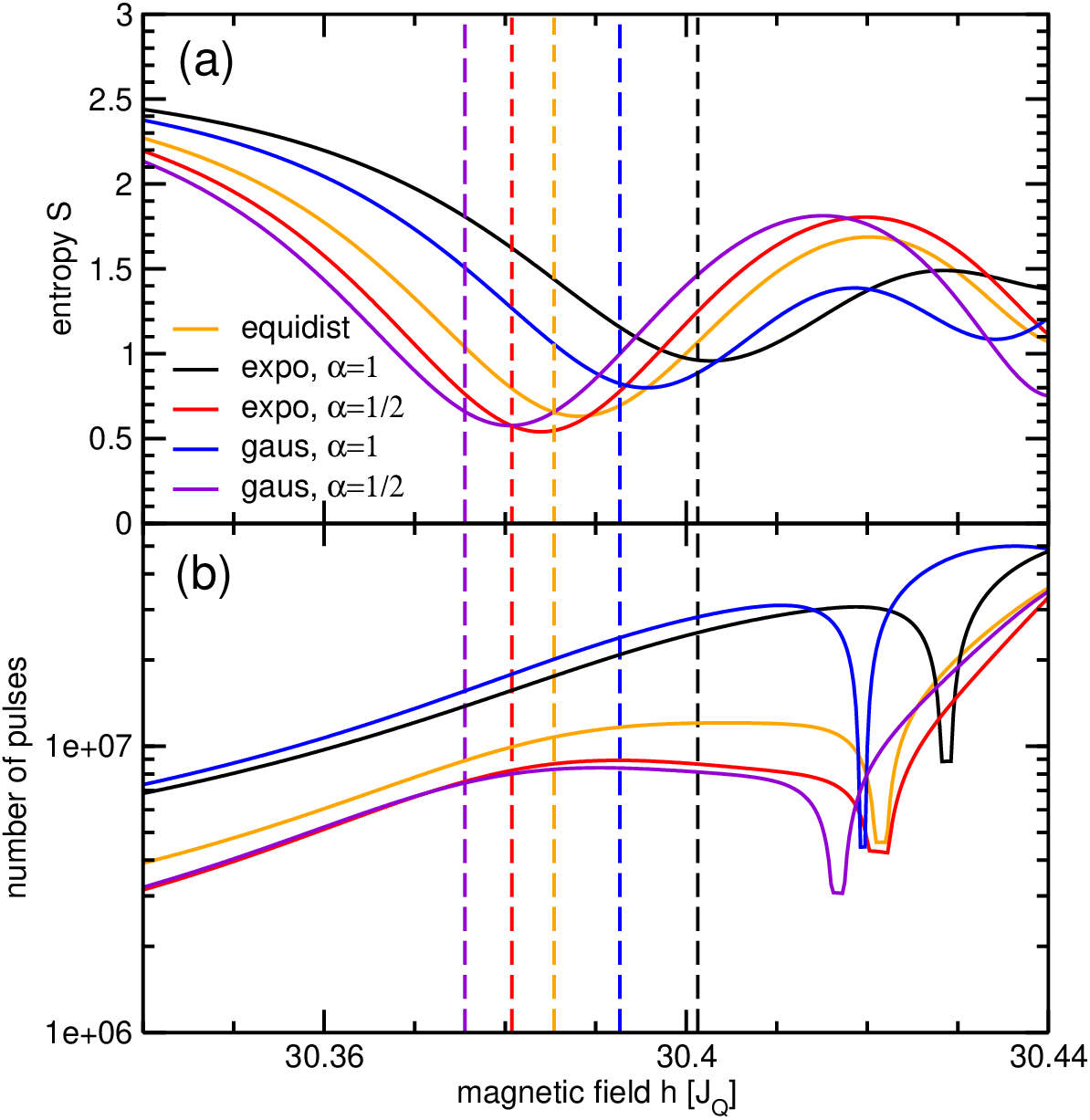} 
\caption{Residual entropies (panel a)  and number of pulses 
(panel b) for a convergence within $1\%$ ($p_\text{thresh}=0.0$)
for $N=3$, $\jm=0.1\jq$, and $z=0.1$ for the equidistant parametrization
in \eqref{eq:equidistant} (legend ``equidist''), the  exponential parametrization
 in \eqref{eq:expo} (legend ``expo'') and the Gaussian parametrization
in \eqref{eq:gaus} (legend ``gaus'').
The vertical dashed lines indicate
the estimates for the entropy minima which are shifted from the
resonances without interactions according to \eqref{eq:over-shift}.}
\label{fig:faster}
\end{figure}

The total numbers of pulses is rather high. As many as $2\cdot 10^{12}$ pulses
for a repetition time $\trp\approx 10$ns imply about six hours of pulsing.
This can be achieved in the lab, but the risk that so far neglected decoherence
mechanisms spoil the process is real. If, however, the pulses can be applied more
frequently, for instance with $\trp=1$ns, the required duration shrinks to about 30 
minutes. The question arises why so many pulses are required. While a comprehensive
study of this aspect is beyond the scope of the present article, first clue
can be given. 

It suggests itself that the slow dynamics in the bath is responsible
for the large number of pulses required for convergence. This idea is corroborated 
by the results displayed in Fig.\ \ref{fig:faster} where a larger
maximum coupling and, importantly, a larger $z$ factor is assumed.
Recall that the $z$-factor is the ratio of the Larmor frequency of the
bath spins to the Larmor frequency of the central spin. If it is increased,
here by a factor of 100, the bath spins precess much quicker.
Indeed, the range of the required number of pulses
is much lower with $2\cdot 10^7$ which is five orders of magnitude less
than for the previous parameters. The former six hours then become fractions
of seconds. Of course, the conventional $g$-factors of nuclear and electronic
spins do not allow for $z=0.1$. But the central spin model as such, built by a central spin
and a bath of spins supplemented by a damped excitation can also be
realized in a different physical system. 

\red{Alternatively, optimized pulses can improve the efficiency of the
distillation by periodic driving. One may either consider modulated pulses
of finite duration \cite{pasin08a} or repeated cycles of several instantaneous
pulses applied at optimized time instants \cite{uhrig07,uhrig07err}
or combinations of both schemes \cite{uhrig10a}. Thus, further research
is called for. The focus, however, of the present work
is to establish the fundamental mechanism built upon periodic driving, dissipation
and commensurability.}

\section{Conclusion}
\label{sec:conclusion}

Previous work has established dynamic nuclear polarization (DNP), for a 
review see Ref.\ \cite{maly08}. But it must be stressed that
the mechanism of this conventional DNP is fundamentally different from
the one described here. Conventionally, the polarization of an electron
is transferred to the nuclear spins, i.e., the polarization of
the electrons induces polarization of the nuclei in the \emph{same} direction.

In contrast, in the setup studied here, the electron is polarized 
in $z$-direction while the nuclear spins are eventually polarized 
perpendicularly in $x$-direction. Hence, the mechanism is fundamentally different:
it is NFF stemming essentially from commensurability. This is
also the distinguishing feature compared to standard optical pumping.
States in the initial mixture which do not allow for a time
evolution commensurate with the repetition time $\trp$ of the pulses are
gradually suppressed \red{while those whose time evolution is 
commensurate are enhanced. This means that the weight of the former
in the density matrix is reduced upon periodic application of the pulses while
the weight of the latter is enhanced. Note that the trace of the
density matrix is conserved so that the suppression of the weight of some states
implies that the weight of other states is increased. The effect of the pulses
on other norms of the density matrix is not obvious since the dynamics is
not unitary, but dissipative.}

\red{For particular magnetic fields, there may be only one particular state allowing
for a  dynamics commensurate with $\trp$. This case leads to the maximum
entropy reduction. Such a}
mechanism can be used also for completely different physical systems,
e.g., in ensembles of oscillators. The studied case of coupled spins
extends the experimental and theoretical observations of NFF for large spin baths 
\cite{greil06a,greil07b,petro12,econo14,beuge16,jasch17,scher18,klein18} where
many values of the polarization of the Overhauser
field can lead to commensurate dynamics. Hence, only a partial reduction of entropy
occurred.

The above established DNP by NFF comprises the potential
for a novel experimental technique for state preparation: laser pulses instead
of microwave pulses as in standard NMR can be employed to prepare
coherent states which can be used for further processing, either to 
perform certain quantum protocols or for analysis of the systems under study.
The combination of optical and radio frequency pulsing appears promising
because it enlarges the possibilities of experimental manipulations.
 Another interesting perspective is to employ the concept of 
state distillation by commensurability to physical systems other than
localized spins, for instance to spin waves in quantum magnets. 
A first experimental observations of commensurability effects 
for spin waves in ferromagnets are already carried out \cite{jackl17}.
\red{Studies on how to enhance the efficiency of the mechanism by
optimization of the shape and distribution of the pulses constitute
an interesting route for further research.}

In summary, we showed that dissipative dynamics of a highly excited
state is sufficient to modify the dynamics of energetically low-lying
spin degrees of freedom away from unitarity. The resulting dynamic map
acts like a contraction converging towards
a single density matrix upon iterated application. The crucial additional 
ingredient is \emph{commensurability} \red{between the external
periodic driving and the internal dynamic processes, for instance Larmor precessions.
If commensurability is possible a substantial entropy reduction
can be induced}, almost to a single pure state. 
This has been explicitly shown for an exemplary small
central spin model including electronic and nuclear Zeeman effect.
This model served as proof-of-principle model to establish the
mechanism of distillation by commensurability.

Such a model describes the electronic spin in quantum dots with 
diluted nuclear spin bath or the spin of unpaired electrons in 
molecules,  hyperfine coupled to nuclear hydrogen spins.
We stress that the mechanism of commensurability can also be put to use in 
other systems with periodic internal processes.
The fascinating potential to create and to manipulate 
coherent quantum states by such approaches deserves further 
investigation.

\section*{Acknowledgements}
The author thanks A.\ Greilich, J.\ Schnack, 
and O.~P.\ Sushkov for useful discussions and the School of Physics of the 
University of New South Wales for its hospitality.

\paragraph{Funding information}
This work was supported by the Deutsche Forschungsgemeinschaft (DFG) 
and the Russian Foundation of Basic Research in TRR 160,
by the DFG in project no.\ UH 90-13/1, and by the Heinrich-Hertz Foundation
of Northrhine-Westfalia.


\begin{appendix}

\section{Derivation of the Linear Mapping}
\label{app:matrix}

The goal is to solve the time evolution of $\rho(t)$ from
just before a pulse until just before the next pulse. 
Since the pulse leads to a unitary time evolution which 
is linear 
\be
\rho(n\trp-)\to\rho(n\trp+)=U_\text{puls}\rho(n\trp-)U_\text{puls}^\dag
\ee
with $U_\text{puls}$ from (5) 
and the subsequent Lindblad dynamics defined by the
linear differential equation (6) 
is linear
as well the total propagation in time is given by 
a linear mapping $M: \rho(n\trp-) \rightarrow \rho((n+1)\trp-)$. 
This mapping is derived here
by an extension of the approach used in Ref.\ \cite{klein18}.

The total density matrix acts on the Hilbert space given by the direct product
of the Hilbert space of the 
central spin comprising three states ($\ket{\uparrow},\ket{\downarrow}, \ket{\text{T}}$)
and the Hilbert space of the spin bath.
We focus on $\rho_\text{TT}:=\bra{\text{T}}\rho\ket{\text{T}}$
which is a $2^N\times2^N$ dimensional density matrix for the spin bath alone because the 
central degree of freedom is traced out. By $\rho_\text{S}$ 
we denote the $d\times d$ dimensional 
density matrix of the spin bath and the central spin, i.e., $d=2^{N+1}$ since
no trion is present: $\rho_\text{S}\ket{\text{T}}=0$. The number of entries
in the density matrix is $D=d^2$, i.e., the mapping we are looking for
can be represented by a $D\times D$ matrix.

The time interval $\trp$ between two  consecutive pulses is sufficiently
long so that all excited trions have decayed before the next pulse arrives.
In numbers, this means $2\gamma\trp\gg 1$ and implies that 
$\rho(n\trp-)=\rho_\text{S}(n\trp-)$ and hence inserting
the unitary of the pulse (5) 
yields
\bs
\label{eq:initial}
\begin{align}
\rho(n\trp+) &= U_\text{puls}\rho_\text{S}(n\trp-)U_\text{puls}^\dag
\\
\label{eq:initial-TT}
\rho_\text{TT}(n\trp+) &=\bra{\uparrow} \rho_\text{S}(n\trp-) \ket{\uparrow}
\\
\rho_\text{S}(n\trp+) &=\ket{\downarrow}\bra{\downarrow} \rho_\text{S}(n\trp-) 
\ket{\downarrow}\bra{\downarrow} \ =\  S^-S^+ \rho_\text{S}(n\trp-) S^-S^+ 
\label{eq:initial-S}
\end{align}
\es
where we used the standard ladder operators $S^\pm$ of the central spin to
express the projection $\ket{\downarrow}\bra{\downarrow}$.
The equations \eqref{eq:initial} set the initial values for the
subsequent Lindbladian dynamics which we derive next.
For completeness, we point out that there are also non-diagonal
contributions of the type $\bra{\text{T}}\rho\ket{\uparrow}$, but they
do not matter for $M$.

Inserting $\rho_\text{TT}$ into the Lindblad equation (6) 
yields 
\be
\label{eq:TT}
\partial_t \rho_\text{TT}(t) = -i [H_\text{nZ},\rho_\text{TT}(t)] -2\gamma \rho_\text{TT}(t).
\ee
No other parts contribute. The solution of \eqref{eq:TT} reads
\be
\label{eq:TT_solution}
\rho_\text{TT}(t) = e^{-2\gamma t} e^{-iH_\text{nZ}t} \rho_\text{TT}(0+)
e^{iH_\text{nZ}t}.
\ee
By the argument $0+$ we denote that the initial density matrix for
the Lindbladian dynamics is the one just after the pulse.

For $\rho_\text{S}$, the Lindblad equation (6) 
implies
\be
\partial_t \rho_\text{S}(t) = -i[H_\text{spin},\rho_\text{S}(t)]
+2\gamma \ket{\uparrow} \rho_\text{TT}(t) \bra{\uparrow}.
\ee
Since we know the last term already from its solution in \eqref{eq:TT_solution}
we can treat it as given inhomogeneity in the otherwise homogeneous
differential equation. With the definition $U_\text{S}(t):= \exp(-iH_\text{spin}t)$
we can write
\be
\partial_t \left(U_\text{S}^\dag(t) \rho_\text{S}(t) U_\text{S}(t)
\right) = 2\gamma U_\text{S}^\dag(t) \ket{\uparrow}\rho_\text{TT}(t)\bra{\uparrow} 
U_\text{S}(t).
\ee
Integration leads to the explicit solution
\be
\rho_\text{S}(t) = U_\text{S}(t) \rho_\text{S}(0+) U_\text{S}^\dag(t)
+2\gamma\int_0^t  U_\text{S}^\dag(t-t') \ket{\uparrow}\rho_\text{TT}(t')\bra{\uparrow} 
U_\text{S}(t-t')dt'.
\label{eq:S_solut}
\ee
If we insert \eqref{eq:TT_solution} into the above equation
we encounter the expression
\be
\ket{\uparrow} \exp(-iH_\text{nZ} t) = \exp(-iH_\text{nZ} t)\ket{\uparrow}
\ =\ \exp(-izh I^x_\text{tot} t) \exp(izh S^x t) \ket{\uparrow}.
\ee
where $I^x_\text{tot} :=S^x+\sum_{i=1}^NI^x_i$ is the total momentum in $x$-direction.
It is a conserved quantity commuting with $H_\text{spin}$ so that a joint
eigenbasis with eigenvalues $m_\alpha$ and $E_\alpha$ exists. We determine 
such a basis $\{\ket{\alpha}\}$ by diagonalization in the
$d$-dimensional Hilbert space ($d=2^{N+1}$) of central spin and spin bath
and convert \eqref{eq:S_solut} in terms 
of the matrix elements of the involved operators. For brevity, we write
$\rho_{\alpha\beta}$ for the matrix elements of $\rho_\text{S}$.
\begin{align}
\rho_{\alpha\beta}(t) &= e^{-i(E_\alpha-E_\beta)t}\Big\{\rho_{\alpha\beta}(0+)
\nonumber \\
&+2\gamma \int_0^t e^{i(E_\alpha-E_\beta-zh(m_\alpha-m_\beta))t'}
\bra{\alpha} e^{izhS^xt'}\ket{\uparrow} \rho_\text{TT}(0+) 
\bra{\uparrow} e^{izhS^xt'}\ket{\beta}dt'\Big\}.
\label{eq:matrix1}
\end{align}

Elementary quantum mechanics tells us that
\be
\label{eq:spin-precess}
e^{izhS^xt'}\ket{\uparrow} =
\frac{1}{2}e^{ia}(\ket{\uparrow}+\ket{\downarrow})
+ \frac{1}{2}e^{-ia}(\ket{\uparrow}-\ket{\downarrow})
\ee
with $a:=zht'/2$ which we need for the last row of equation \eqref{eq:matrix1}.
Replacing $\rho_\text{TT}(0+)$ by 
$\bra{\uparrow} \rho_\text{S}(n\trp-) \ket{\uparrow}$ according to \eqref{eq:initial-TT}
and inserting \eqref{eq:spin-precess} we obtain
\bs
\begin{align}
 \bra{\alpha} e^{izhS^xt'}\ket{\uparrow} \rho_\text{TT}(0+) 
\bra{\uparrow} e^{izhS^xt'}\ket{\beta} 
 & =\bra{\alpha} e^{izhS^xt'}\ket{\uparrow} \bra{\uparrow} \rho_\text{S}(0-) \ket{\uparrow}
\bra{\uparrow} e^{izhS^xt'}\ket{\beta}
\\
& = \frac{1}{2} \left( R^{(0)} + e^{izht'} R^{(1)} + e^{-izht'} R^{(-1)}
\right)_{\alpha\beta}
\label{eq:spin2}
\end{align}
\es
with the three $d\times d$ matrices
\bs
\begin{align}
R^{(0)} &:= S^+S^- \rho_\text{S}(0-) S^+S^- + S^- \rho_\text{S}(0-) S^+
\\
R^{(1)} &:= \frac{1}{2}(S^++\id_d)S^- \rho_\text{S}(0-) S^+(S^--\id_d)
\\
R^{(-1)} &:= \frac{1}{2}(S^+-\id_d)S^- \rho_\text{S}(0-) S^+(S^-+\id_d).
\end{align}
\es
In this derivation, we expressed ket-bra combinations 
by the spin ladder operators according to
\be
\ket{\uparrow} \bra{\uparrow} = S^+S^- \qquad
 \ket{\uparrow} \bra{\downarrow} = S^+ \qquad
 \ket{\downarrow} \bra{\uparrow} = S^-.
\ee

The final step consists in inserting \eqref{eq:spin2} into \eqref{eq:matrix1}
and integrating the exponential time dependence straightforwardly from 0 to $\trp$. 
Since we assume that $2\gamma\trp\gg1$ so that no trions are present once the next
pulse arrives the upper integration limit $\trp$ can safely and consistently be
replaced by $\infty$. This makes the expressions
\be
G_{\alpha\beta}(\tau) := \frac{\gamma}{2\gamma-i[E_\alpha-E_\beta+zh(m_\beta-m_\alpha+\tau)]}
\ee
appear where $\tau\in\{-1,0,1\}$. Finally,
we use \eqref{eq:initial-S} and summarize
\be
\rho_{\alpha\beta}(t) = e^{-i(E_\alpha-E_\beta)t}
\Big\{
(S^-S^+ \rho_\text{S}(0-) S^-S^+ )_{\alpha\beta}
 +\sum_{\tau=-1}^1 G_{\alpha\beta}(\tau) R^{(\tau)}_{\alpha\beta}
\Big\}.
\label{eq:complete}
\ee
This provides the complete solution for the dynamics of $d\times d$ matrix
$\rho_\text{S}$ 
from just before a pulse ($t=0-$) till just before the next pulse for
which we set $t=\trp$ in \eqref{eq:complete}.

In order to set up the linear mapping $M$ as $D\times D$ dimensional matrix
with $D=d^2$ we denote the matrix elements $M_{\mu'\mu}$ where $\mu$
is a combined index for the index pair $\alpha\beta$ and $\mu'$
for $\alpha'\beta'$  with 
$\alpha,\beta,\alpha',\beta'\in\{1,2\ldots,d\}$. 
For brevity, we introduce
\be
P_{\alpha\beta} := [(S^++\id_d)S^-]_{\alpha\beta} \qquad
Q_{\alpha\beta} := [(S^+-\id_d)S^-]_{\alpha\beta}.
\ee
Then, \eqref{eq:complete} implies
\begin{align}
M_{\mu'\mu} &= \frac{1}{2}e^{-i(E_{\alpha'}-E_{\beta'})\trp}\Big\{
2(S^-S^+)_{\alpha'\alpha} (S^-S^+)_{\beta\beta'}
\nonumber \\
&+2G_{\alpha'\beta'}(0)
\left[(S^+S^-)_{\alpha'\alpha} (S^+S^-)_{\beta\beta'}
+ S^-_{\alpha'\alpha} S^+_{\beta\beta'}\right]
\nonumber \\
&+\left[
G_{\alpha'\beta'}(1) P_{\alpha'\alpha} Q^*_{\beta'\beta}
+ G_{\alpha'\beta'}(-1) Q_{\alpha'\alpha} P^*_{\beta'\beta}
\right]\Big\}.
\label{eq:matrix2}
\end{align}
This concludes the explicit derivation of the matrix elements
of $M$. Note that they are relatively simple in the sense that no
sums over matrix indices are required on the right hand side of
\eqref{eq:matrix2}. This relative simplicity is achieved because
we chose to work in the eigenbasis of $H_\text{spin}$.
Other choices of basis are possible, but render the explicit
respresentation significantly more complicated.


\section{Properties of the Time Evolution}
\label{app:properties}

\paragraph{Preliminaries}
Here we state several mathematical properties of the mapping $M$ which
hold for any Lindblad dynamics over a given time interval which can
be iterated arbitrarily many times. We assume
that the underlying Hilbert space is $d$ dimensional so that $M$ acts
on the $D=d^2$ dimensional Hilbert space of 
$d\times d$ matrices, i.e., $M$ can be seen as $D\times D$ matrix. 
We denote the standard scalar product in the space of operators by
\be
(A|B):=\tr(A^\dag B)
\ee
where the trace refers to the $d\times d$ matrices $A$ and $B$.

Since no state of the physical system vanishes in its temporal evolution
$M$ conserves the trace of any density matrix 
\be
\tr(M\rho)=\tr(\rho).
\ee
This implies that $M$ conserves the trace of \emph{any} operator $C$. This can be
seen by writing $C=(C+C^\dag)/2+ (C-C^\dag)/2=R+iG$ where $R$ and $G$ are
hermitian operators. They can be diagonalized and split into their positive and their
negative part $R=p_1-p_2$ and $G=p_3-p_4$. Hence, each $p_i$ is a density matrix
up to some real, positive scaling and we have
\be
\label{eq:C-darst}
C = p_1-p_2+i(p_3-p_4).
\ee
Then we conclude
\bs
\begin{align}
\tr(MC) &= \tr(Mp_1)-\tr(Mp_2)+ i(\tr(Mp_3)-\tr(Mp_4))
\\
& =  \tr(p_1)-\tr(p_2)+i(\tr(p_3)-\tr(p_4))
\ =\ \tr(C).
\end{align}
\es

\paragraph{Property 1.}
The conservation of the trace for any $C$ implies 
\be
\label{eq:gen_trace_conserv}
\tr(C)=(\id_d|C)=(\id_d|MC)=(M^\dag\id_d|C)
\ee
where $\id_d$ is the $d\times d$-dimensional identity matrix and $M^\dag$ is the 
$D\times D$ hermitian conjugate of $M$. From \eqref{eq:gen_trace_conserv} we conclude
\be
M^\dag \id_d = \id_d
\ee
which means that $\id_d$ is an eigenoperator of $M^\dag$ with eigenvalue 1. Since the 
characteristic polynomial of $M$ is the same as the one of $M^\dag$ up to complex conjugation
we immediately see that $1$ is also an eigenvalue of $M$.
If the dynamics of the system takes place in $n$ independent subspaces without
transitions between them, the $n$ different 
traces over these subspaces are conserved separately
Such a separation occurs in case conserved symmetries
split the Hilbert space, for instance
the total spin is conserved in the dynamics given by 
(6) 
if all couplings are equal. 
Then, the above argument implies the existence of $n$ different eigenoperators with
eigenvalue 1.  Hence the degeneracy is (at least) $n$ which proves property
1. in the main text.

\paragraph{Properties 2. and 3.} As for property 2, we consider an eigenoperator $C$ of 
$M$ with eigenvalue $\lambda\neq1$ so that $MC=\lambda C$. Then
\be
\tr(C) = \tr(MC) \ = \ \lambda \tr(C)
\ee
implies $\tr(C)=0$, i.e., tracelessness as stated.
Since all density matrices can be written as linear combinations of eigenoperators
there must be at least one eigenoperator with finite trace. 
In view of property 2., this needs to be an
eigenoperator with eigenvalue 1 proving property 3.
The latter conclusion holds true if we assume that $M$ cannot be 
diagonalized, but only has a Jordan normal form. If $d_\text{J}$ is the 
dimension of the largest Jordan block, the density matrix $M^{d_\text{J}-1}\rho$
will be a linear combination of eigenoperators while still having the trace 1.

\paragraph{Property 4.} Next, we show that no eigenvalue $\lambda$ can be larger than 1 in absolute value.
The idea of the derivation is that the iterated application
of $M$ to the eigenoperator belonging to $|\lambda|>1$ would  make
this term grow exponentially $\propto |\lambda|^n$
beyond any bound which cannot be true. The formal proof is a bit intricate.

First, we state that for any two density matrices $\rho$ and $\rho'$
their scalar product is non-negative $0\le (\rho|\rho')$ because it
can be viewed as expectation value of one of them with respect to the other 
and both are positive operators. In addition, the Cauchy-Schwarz inequality
implies
\be
\label{eq:rhorho}
0 \le (\rho|\rho') \le \sqrt{(\rho|\rho)(\rho'|\rho')}
\ =\ \sqrt{\tr(\rho^2)\tr((\rho')^2)} \ \le\  1.
\ee

Let $C$ be the eigenoperator of $M^\dag$ belonging to $\lambda$; it may be represented
as in \eqref{eq:C-darst} and scaled such that the maximum of the traces of the $p_i$ is
1. Without loss of generality this is the case for $p_1$, i.e., $\tr(p_1)=1$. Otherwise, 
$C$ is simply rescaled: by $C\to-C$ to switch $p_2$ to $p_1$, by $C\to-iC$
to switch $p_3$ to $p_1$, or by $C\to iC$ to switch $p_4$ to $p_1$.
On the one hand, we have for any density matrix $\rho_n$
\be
|(C|\rho_n)| \le |\Re(C|\rho_n)| + |\Im(C|\rho_n)| \le 2
\ee
where the last inequality results form \eqref{eq:rhorho}.
On the other hand, we set $\rho_n:= M^np_1$ and obtain
\bs
\begin{align}
2 &\ge |(C|\rho_n)|
 = |((M^\dag)^nC|p_1)|
= |\lambda^*|^n |(C|p_1)|
 =|\lambda|^n \sqrt{(\Re(C|p_1))^2+ (\Im(C|p_1))^2}
\\
&\ge |\lambda|^n|\Re(C|p_1)|
 = |\lambda|^n(p_1|p_1)
\end{align}
\es
where we used $(p_1|p_2)=0$ in the last step; this holds because
$p_1$ and $p_2$ result from the same diagonalization, but refer
to eigenspaces with eigenvalues of different sign.
In essence we derived
\be
2 \ge |\lambda|^n(p_1|p_1)
\ee
which clearly implies a contradiction for $n \to \infty$ because
the right hand side increases to infinity for $|\lambda|>1$.
Hence there cannot be eigenvalues with modulus larger than 1.

\paragraph{Property 5.} The matrix $M$ can be represented 
with respect to a basis of the  Krylov space spanned by the 
operators 
\be
\label{eq:krylov}
\rho_n:=M^n\rho_0
\ee
 where $\rho_0$ is an arbitrary initial density matrix which should
contain contributions from all eigenspaces of $M$. For instance,
a Gram-Schmidt algorithm applied to the Krylov basis generates
an orthonormal basis $\tilde \rho_n$. Due to the fact, that all the 
operators $\rho_n$ from \eqref{eq:krylov} are hermitian density matrices 
$\tilde \rho_n = \tilde \rho_n^\dag$, we know that
all overlaps $(\rho_m|\rho_n)$ are real and hence the constructed 
orthonormal basis $\tilde \rho_n$ consists of hermitian operators.
Also, all matrix elements
 $(\rho_m|M\rho_n)=(\rho_m|\rho_{n+1})$ are real so that the resulting 
representation $\tilde M$ is a matrix with real coefficients whence 
\bs
\be
\tilde M c = \lambda c
\ee
implies 
\be
\tilde M c^* = \lambda^* c^*
\ee
\es
by complex conjugation. Here $c$ is a vector of complex numbers $c_n$ which
define the corresponding eigenoperators by
\be
\label{eq:cC}
C = \sum_{n=1}^D c_n \tilde \rho_n.
\ee
Thus, $c$ and $c^*$ define $C$ and $C^\dag$, respectively.

\paragraph{Property 6.} In view of the real representation $\tilde M$ of $M$
with respect to an orthonormal basis of hermitian operators derived in the
previous paragraph the determination of the eigenoperators with eigenvalue
1 requires the computation of the kernel of $\tilde M-\id_D$. This is a linear algebra
problem in $\mathbb{R}^D$ with real solutions which correspond to
hermitian operators by means of \eqref{eq:cC}. This shows the stated
property 6..

\begin{figure}[hbt]
\centering
\includegraphics[width=0.49\columnwidth]{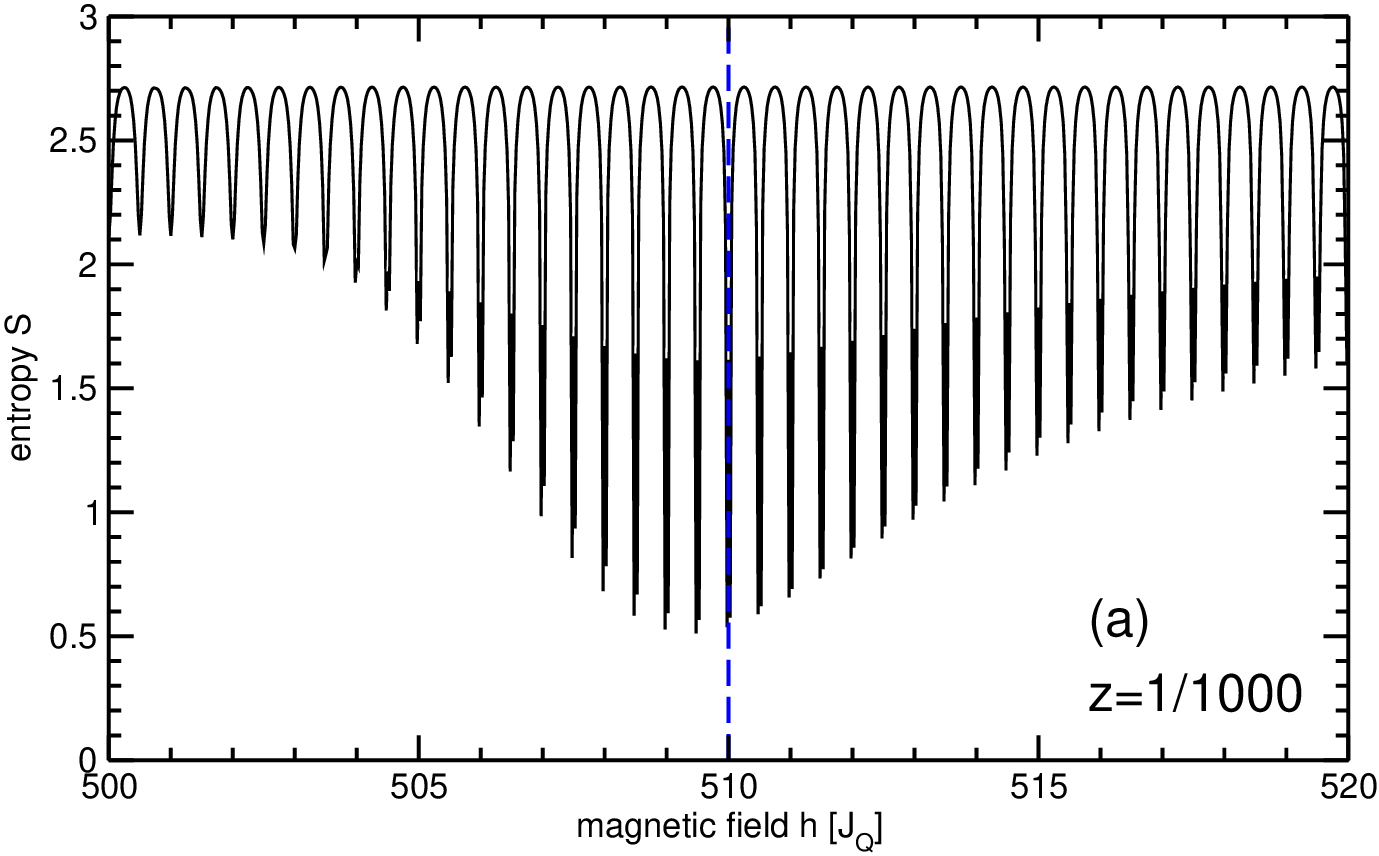}
\includegraphics[width=0.49\columnwidth]{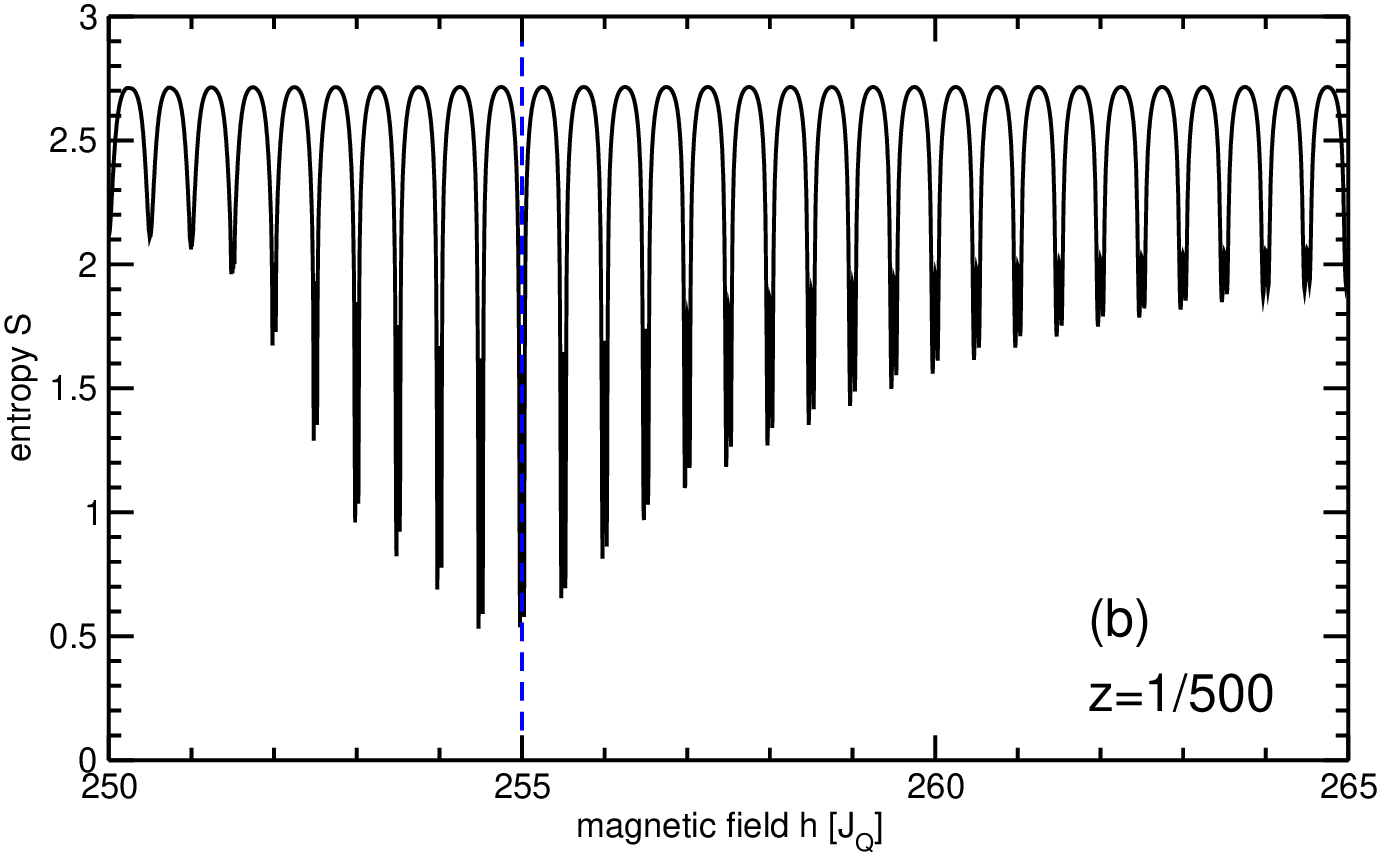} 
\includegraphics[width=0.49\columnwidth]{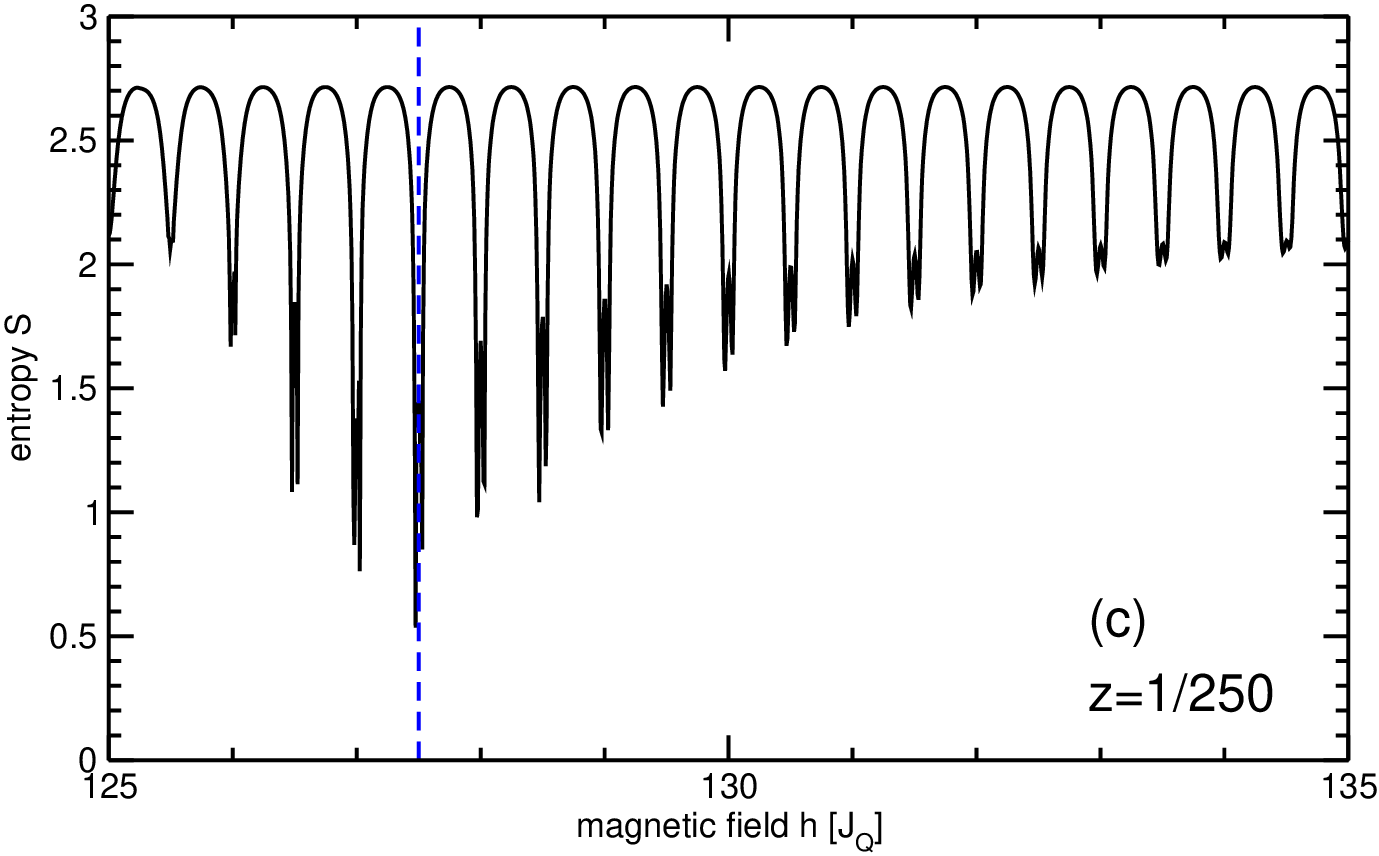} 
\caption{(a) Residual entropy as function of the applied
magnetic field for $N=3, \jm=0.02$, and $z=1/1000$ to show the  position 
at $h=2\pi/(z\trp)$ and the shift, dashed line at $\approx 500\jq\jm/(2z)$ 
of the nuclear magnetic resonance. 
(b) Same as (a) for $z=1/500$.  (c) Same as (a) for $z=1/250$.}
\label{fig:z-depend}
\end{figure}


\section{Shift of the Nuclear Resonance}
\label{app:shift}

\begin{figure}[hbt]
\centering
\includegraphics[width=0.5\columnwidth]{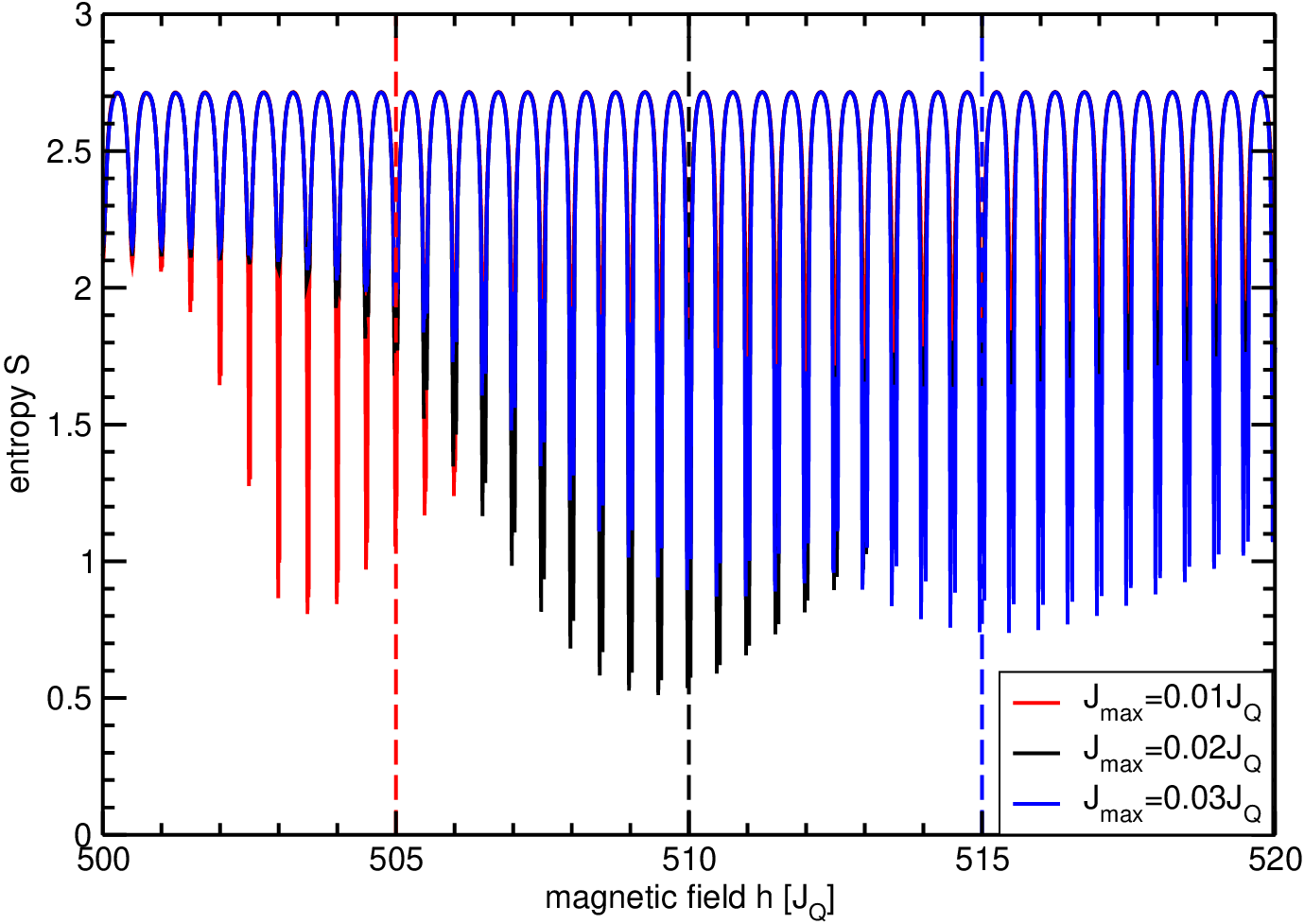}
\caption{Residual entropy as function of the applied
magnetic field for $N=3,  z=1/1000$ and various values of $\jm$.
The shifts indicated by the dashed lines 
correspond to the estimate \eqref{eq:nuclear-shift2}.
\label{fig:jm-depend}}
\end{figure}

In the main text, the shift of the nuclear resonance due to the 
coupling of the nuclear spins to the central, electronic spin
was shown in the right panel of Fig.\ 
1(a).  The effect can be estimated by 
\be
\label{eq:nuclear-shift2}
z\Delta h \approx \pm \jm/2.
\ee
This relation is highly plausible, but it cannot be derived analytically
because no indication for a polarization of the central, electronic
spin in $x$-direction was found. Yet, the numerical data
corroborates the validity of \eqref{eq:nuclear-shift2}.

In Fig.\ \ref{fig:z-depend}, we show that the nuclear resonance
without shift occurs for
\be
zh\trp=2\pi n'
\ee
where $n'\in\mathbb{Z}$. But it is obvious that
an additional shift occurs which is indeed captured by 
\eqref{eq:nuclear-shift2}.

In order to support \eqref{eq:nuclear-shift2} further, we also 
study various values of $\jm$ in Fig.\ \ref{fig:jm-depend}.
The estimate \eqref{eq:nuclear-shift2} captures the main trend
of the data, but it is not completely quantitative because the
position of the dashed lines relative to the minimum of the
envelope of the resonances varies slightly for different values of $\jm$.
Hence, a more quantitative explanation is still called for. 


\section{Entropy Reduction for Other Distributions of Couplings}
\label{app:other}

In the main text, we analyzed a uniform distribution of couplings, see
Eq.\ (8).
In order to underline that our results are generic and not linked to 
a special distribution, we provide additional results for
two distributions which are often considered in literature, namely an
exponential parameterization as defined in \eqref{eq:expo}
 and a Gaussian parametrization as defined in \eqref{eq:gaus}.

\begin{figure}[htb]
\centering
\includegraphics[width=0.49\columnwidth]{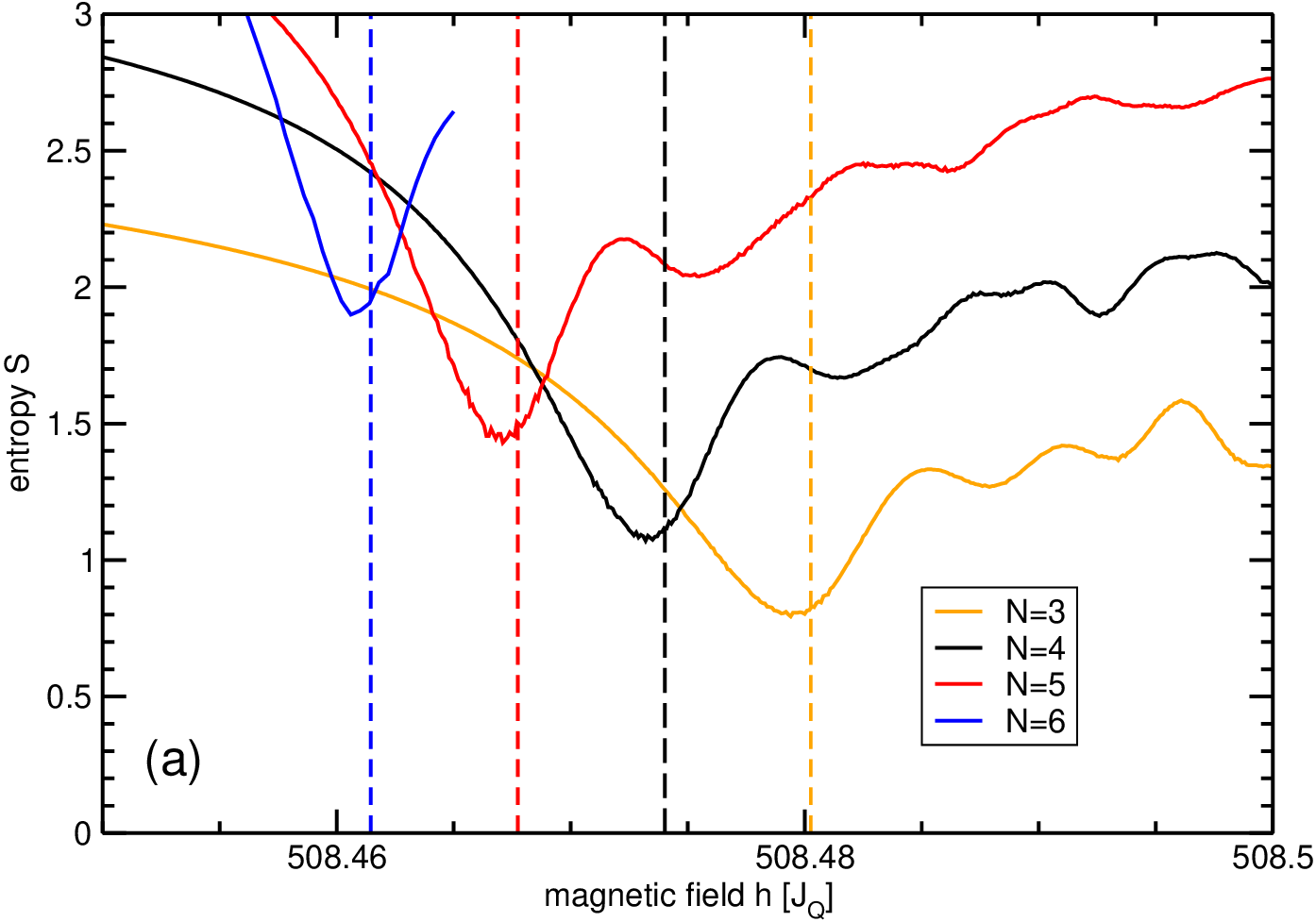}
\includegraphics[width=0.48\columnwidth]{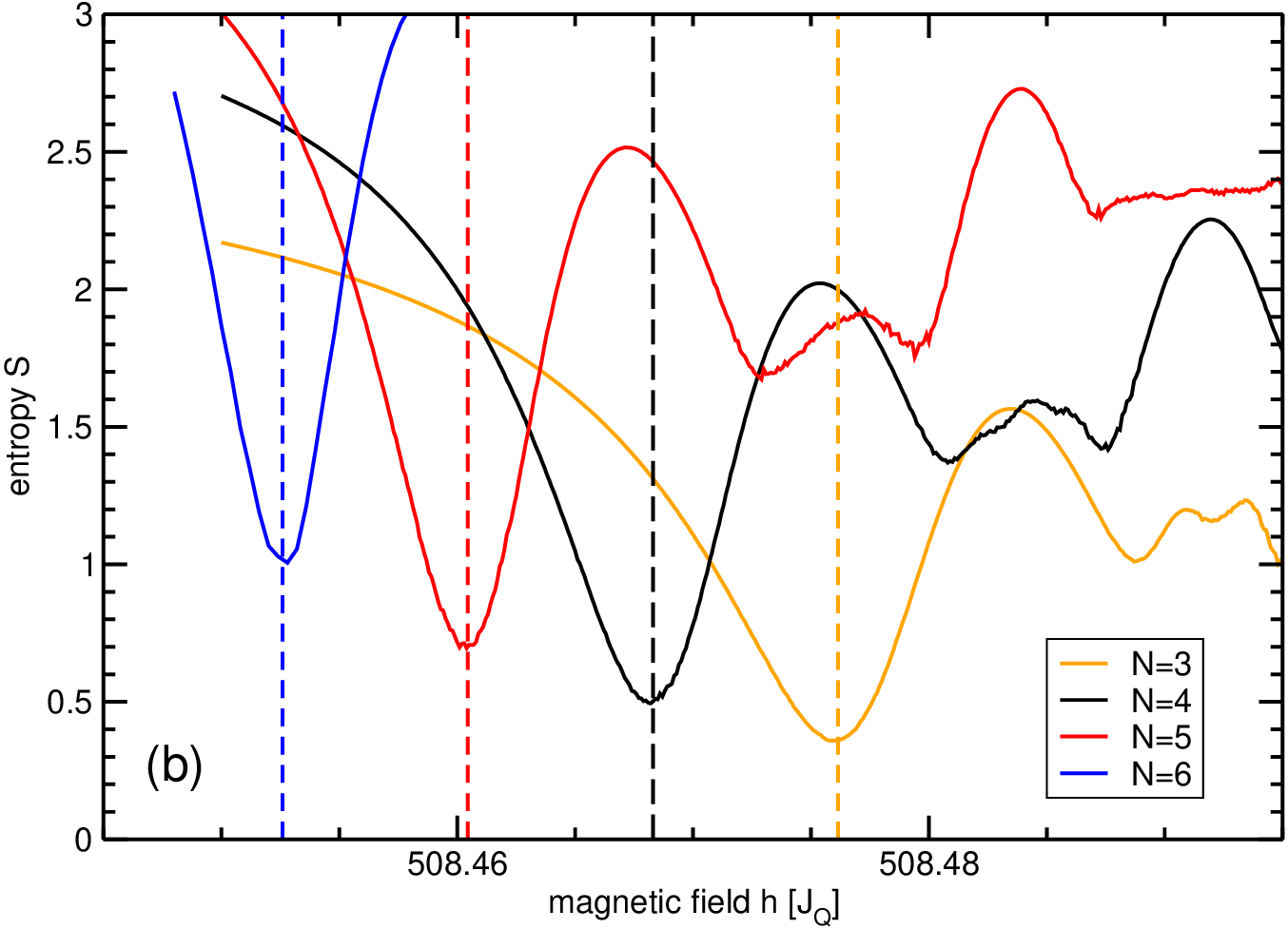}
\caption{Residual entropy as function of the applied
magnetic field for various bath sizes $N$ for
the exponentially distributed couplings given by \eqref{eq:expo}; panel (a) for
$\alpha=1$ and panel (b) for $\alpha=0.5$ and hence smaller ratio
$J_\text{min}/J_\text{max}$.
\label{fig:expo}}
\end{figure}

The key difference between both parametrizations \eqref{eq:expo}
and \eqref{eq:gaus} is that due to the quadratic argument in \eqref{eq:gaus}
the large couplings in this parametrization are very close to each other,
in particular for increasing $N$. Hence, one can study whether this feature
is favorable of unfavorable for entropy reduction.

\begin{figure}[htb]
\centering
\includegraphics[width=0.49\columnwidth]{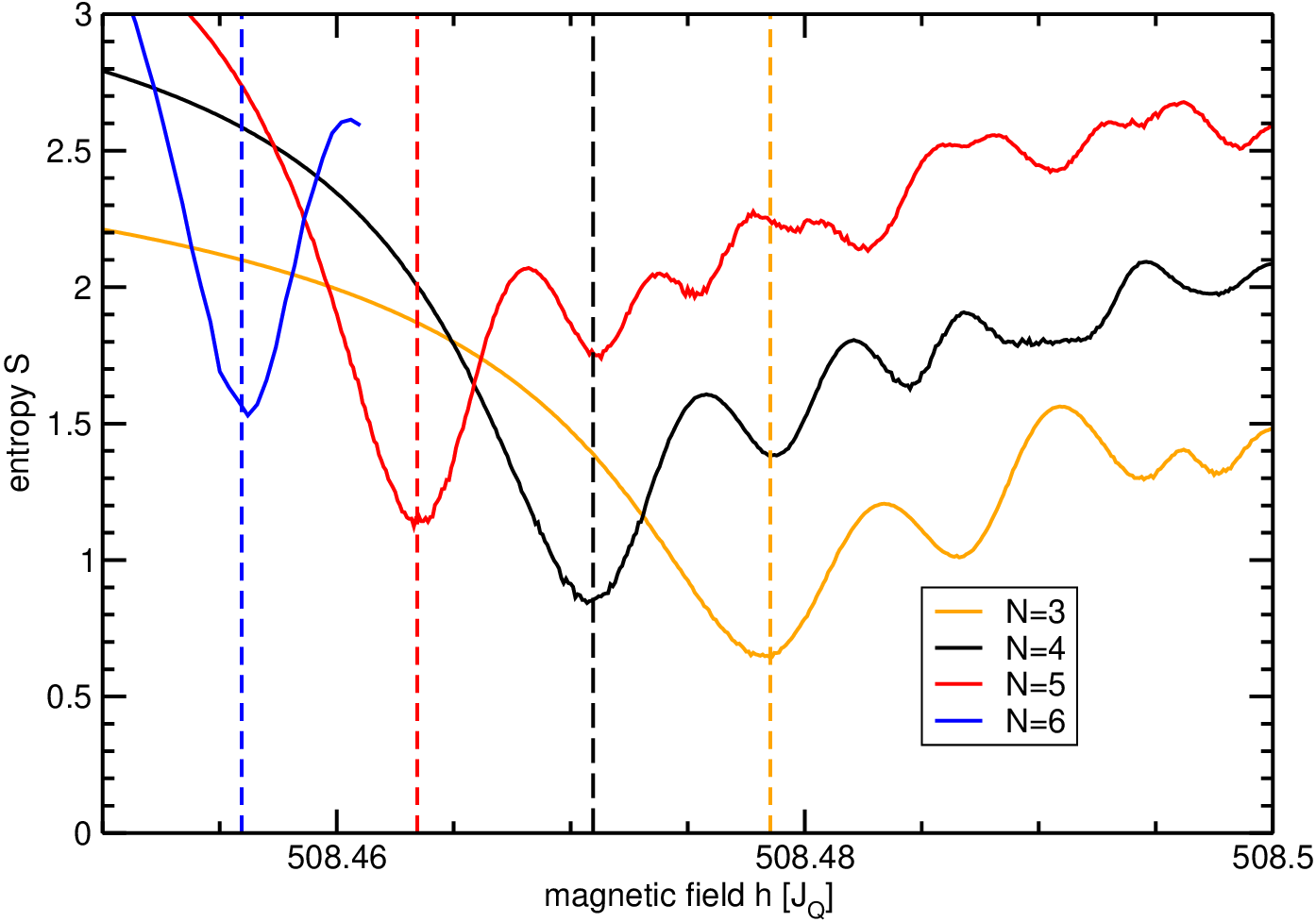}
\includegraphics[width=0.48\columnwidth]{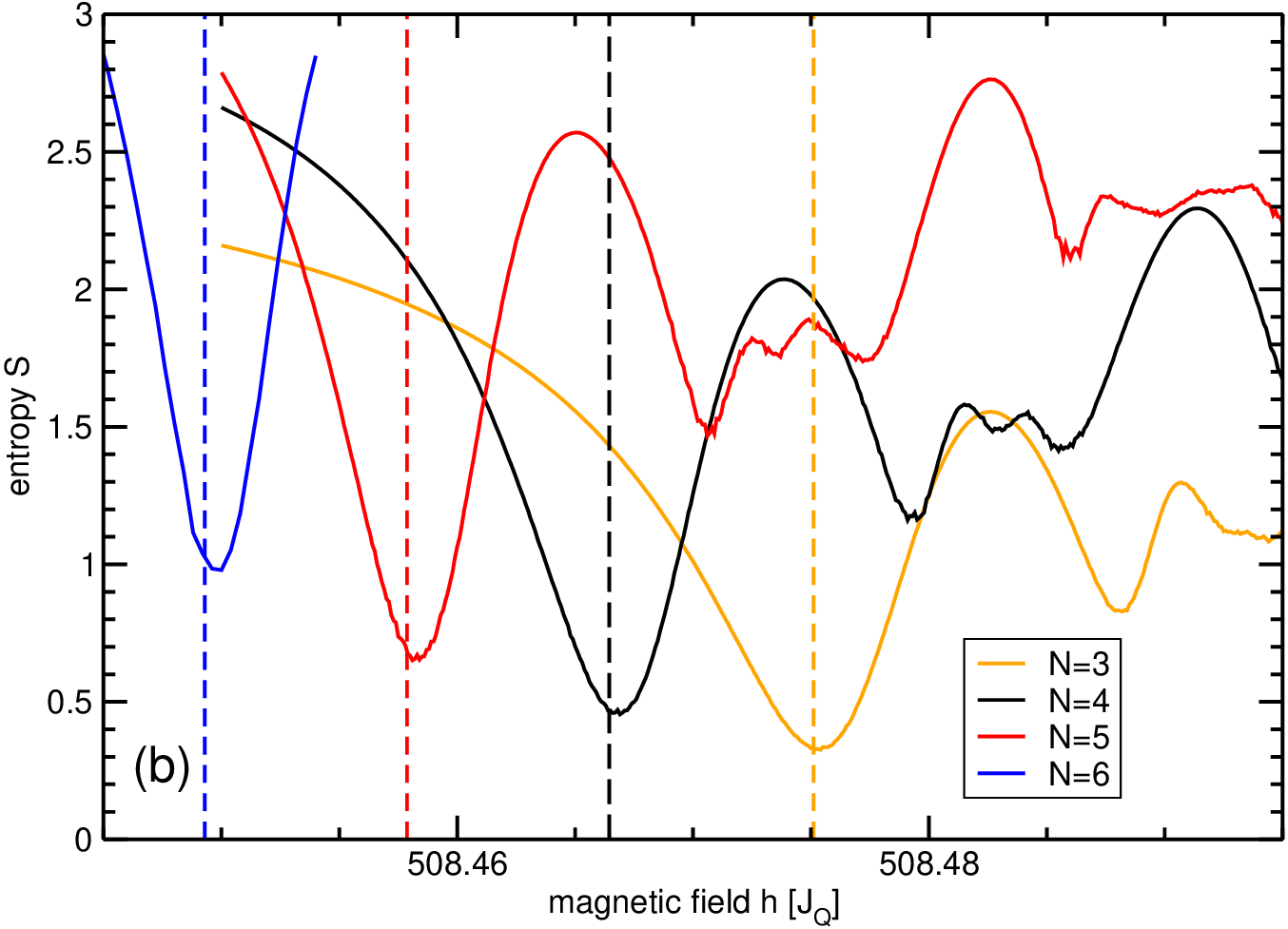}
\caption{Residual entropy as function of the applied
magnetic field for various bath sizes $N$ for
the Gaussian distributed couplings given by \eqref{eq:gaus}; panel (a) for
$\alpha=1$ and panel (b) for $\alpha=0.5$ and hence smaller ratio
$J_\text{min}/J_\text{max}$.
\label{fig:gaus}}
\end{figure}

Additionally, the difference between $\alpha=0.5$ and $\alpha=1$ consists
in a different spread of the couplings. For $\alpha=1$, one has
$J_\text{min}/J_\text{max}=1/e$ in both parametrizations while one has
$J_\text{min}/J_\text{max}=1/\sqrt{e}$ for $\alpha=0.5$, i.e., the spread is
smaller.

Figure \ref{fig:expo} displays the results for the exponential parametrization
\eqref{eq:expo} while Fig.\ \ref{fig:gaus} depicts the results for the Gaussian
parametrization \eqref{eq:gaus}. Comparing both figures shows that the
precise distribution of the couplings does not matter much. Exponential and 
Gaussian parametrization lead to very similar results. They also strongly ressemble
the results shown in Fig.\ 2a in the main text for a uniform
distribution of couplings. This is quite remarkable since the Gaussian
parametrization leads to couplings which are very close to each other
and to the maximum coupling. This effect does not appear to influence
the achievable entropy reduction. 

The ratio $J_\text{min}/J_\text{max}$ between the smallest to the largest coupling
appears to have an impact. If it is closer to unity, here for
$\alpha=0.5$, the reduction of entropy works even better than 
for smaller ratios.
\end{appendix}


\nolinenumbers

\end{document}